\documentclass[runningheads]{llncs}

\usepackage{eccv}

\usepackage{eccvabbrv}

\usepackage{graphicx}
\usepackage{booktabs}
\usepackage{algorithmic}
\usepackage{algorithm}

\usepackage[accsupp]{axessibility}  

\usepackage[pagebackref,breaklinks,colorlinks,citecolor=eccvblue]{hyperref}

\usepackage{orcidlink}
\usepackage{colortbl}
\PassOptionsToPackage{prologue,dvipsnames}{xcolor}

\usepackage{amssymb} 
\usepackage{amsmath}
\usepackage{array}
\usepackage{multirow}
\definecolor{lightyellow}{rgb}{1,0.98,0.8}
\definecolor{lightred}{rgb}{1,0.8,0.8}

\begin{document}

\title{GaussianImage: 1000 FPS Image Representation and Compression by 2D Gaussian Splatting} 

\titlerunning{GaussianImage}

\author{
Xinjie Zhang\inst{1,3}\thanks{Equal Contribution. $^\dotplus$ This work was done when Xinjie Zhang and Xingtong Ge interned at SenseTime Research. $^{\dagger}$ Corresponding Authors.}$^\dotplus$ \and
Xingtong Ge\inst{2,3}$^{\star \dotplus}$ \and
Tongda Xu\inst{4} \and
Dailan He\inst{5} \and \\
Yan Wang\inst{4} \and
Hongwei Qin\inst{3} \and
Guo Lu\inst{6} \and
Jing Geng$^{2 \dagger}$ \and
Jun Zhang$^{1 \dagger}$
}

\authorrunning{Xinjie Zhang, Xingtong Ge et al.}


\institute{$^{1}$ The Hong Kong University of Science and Technology \\ $^{2}$ Beijing Institute of Technology \quad $^{3}$ SenseTime Research \\ 
$^{4}$ Institute for AI Industry Research (AIR), Tsinghua University \\$^{5}$ The Chinese University of Hong Kong \quad $^{6}$ Shanghai Jiaotong University \\
\email{\scriptsize xzhangga@connect.ust.hk, xingtong.ge@gmail.com, x.tongda@nyu.edu \\ hedailan@link.cuhk.edu.hk, wangyan202199@163.com, qinhongwei@sensetime.com \\ luguo2014@sjtu.edu.cn, janegeng@bit.edu.cn, eejzhang@ust.hk}
}


\maketitle

\begin{abstract}
Implicit neural representations (INRs) recently achieved great success in image representation and compression, offering high visual quality and fast rendering speeds with 10-1000 FPS, assuming sufficient GPU resources are available. However, this requirement often hinders their use on low-end devices with limited memory. 
In response, we propose a groundbreaking paradigm of image representation and compression by 2D Gaussian Splatting, named GaussianImage. We first introduce 2D Gaussian to represent the image, where each Gaussian has 8 parameters including position, covariance and color. Subsequently, we unveil a novel rendering algorithm based on accumulated summation. Remarkably, our method with a minimum of 3$\times$ lower GPU memory usage and 5$\times$  faster fitting time not only rivals INRs (e.g., WIRE, I-NGP) in representation performance, but also delivers a faster rendering speed of 1500-2000 FPS regardless of parameter size. Furthermore, we integrate existing vector quantization technique to build an image codec. Experimental results demonstrate that our codec attains rate-distortion performance comparable to compression-based INRs such as COIN and COIN++, while facilitating decoding speeds of approximately 2000 FPS. Additionally, preliminary proof of concept shows that our codec surpasses COIN and COIN++ in performance when using partial bits-back coding. Code is available at \url{https://github.com/Xinjie-Q/GaussianImage}.

\end{abstract}

\section{Introduction}
\label{sec:intro}
Image representation is a fundamental tasks in signal processing and computer vision. Traditional image representation methods, including grid graphics, wavelet transform \cite{heil1989continuous}, and discrete cosine transform \cite{ahmed1974discrete}, have been extensively applied across a broad spectrum of applications, from image compression to vision task analysis. However, these techniques encounter significant obstacles when processing large-scale datasets and striving for highly efficient storage solutions.


The advent of implicit neural representations (INRs) \cite{stanley2007compositional, sitzmann2020implicit} marks a significant paradigm shift in image representation techniques. Typically, INRs employ a compact neural network to derive an implicit continuous mapping from input coordinates to the corresponding output values. This allows INRs to capture and retain image details with greater efficiency, which provides considerable benefits across various applications, including image compression \cite{dupont2021coin, dupont2022coin++, strumpler2022implicit, guo2024compression}, deblurring \cite{xu2022signal, linr2023, quan2023single}, and super-resolution \cite{chen2021learning, ma2022recovering, nguyen2023single}. However, most state-of-the-art INR methods \cite{sitzmann2020implicit, fathony2020multiplicative, tancik2020fourier, ramasinghe2022beyond, saragadam2023wire} rely on a large high-dimensional multi-layer perceptron (MLP) network to accurately represent high-resolution images. This dependency leads to prolonged training times, increased GPU memory requirements, and slow decoding speed. While recent innovations \cite{martel2021acorn, takikawa2021neural, muller2022instant, chen2023neurbf} have introduced multi-resolution feature grids coupled with a compact MLP to accelerate training and inference, they still require enough GPU memory to support their fast training and inference, which is difficult to meet when resources are limited. Consequently, these challenges substantially hinder the practical deployment of INRs in real-world scenarios.

In light of these challenges, our research aims to develop an advanced image representation technique that enables efficient training, friendly GPU memory usage, and fast decoding. To achieve this goal, we resort to Gaussian Splatting (GS) \cite{kerbl20233d} that is recently developed for 3D scene reconstruction. By leveraging explicit 3D Gaussian representations and differentiable tile-based rasterization, 3D GS not only enjoys high visual quality with competitive training times, but also achieves real-time rendering capabilities. 

\begin{figure*}[t]
  \centering
  \subfloat
  {\includegraphics[scale=0.4]{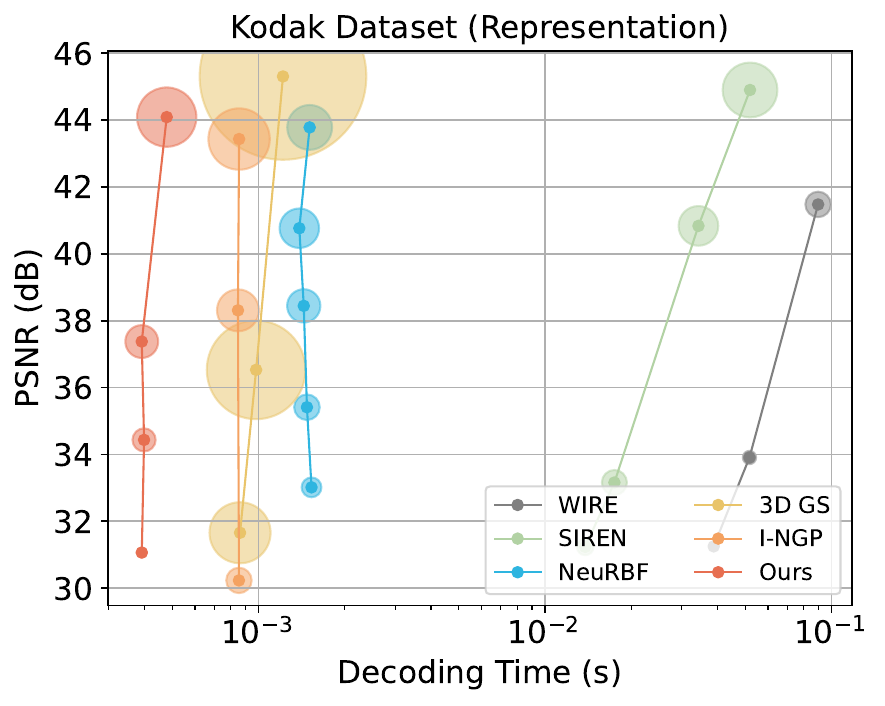}}
  \subfloat
  {\includegraphics[scale=0.4]{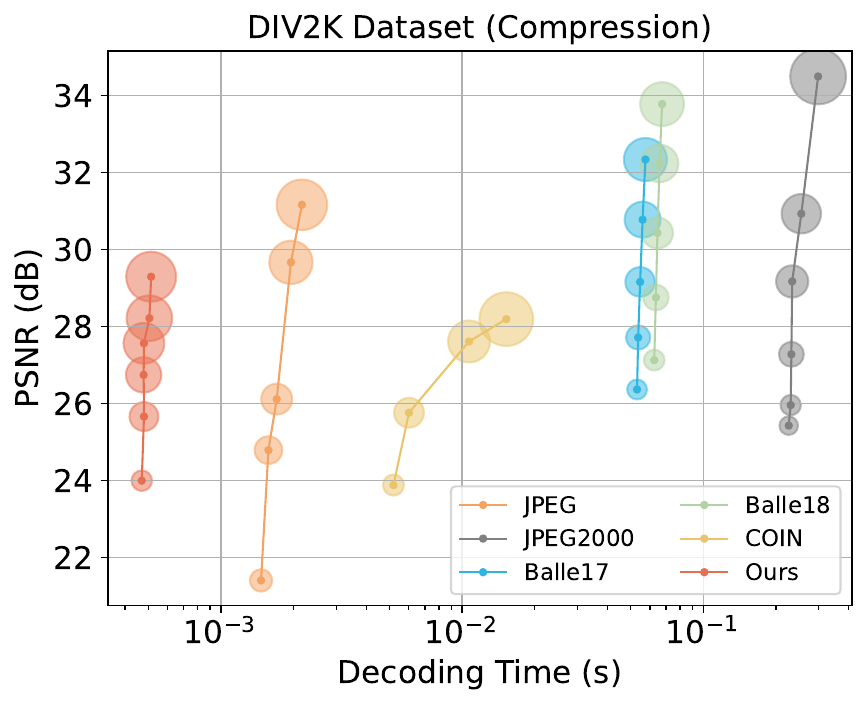}}
  \vspace{-0.2cm}
  \caption{Image representation (left) and compression (right) results with different decoding time on the Kodak and DIV2K dataset, respectively. The radius of each point indicates the parameter size (left) or bits per pixel (right). Our method enjoys the fastest decoding speed regardless of parameter size or bpp.}
  \label{fig:image_representation}
  \vspace{-0.4cm}
\end{figure*}

Nevertheless, it is non-trivial to directly adapt 3D GS for efficient single image representation. \textbf{Firstly}, considering that existing 3D GS methods \cite{kerbl20233d, chen2024survey} depend on varying camera transformation matrices to render images from different perspectives, a straightforward adaptation for single image is fixing the camera transformation matrix to render an image from a single viewing angle. Unfortunately, each 3D Gaussian usually includes 59 learnable parameters \cite{kerbl20233d} and thousands of 3D Gaussians are required for representing a single image. This naive approach substantially increases the storage and communication demands. As can be inferred from \tableautorefname~\ref{table:image_representation}, the storage footprint for a single image with tens of kilobytes can escalate to dozens of megabytes, which makes rendering difficult on low-end devices with limited memory. 
\textbf{Secondly}, the rasterization algorithm \cite{kerbl20233d} in 3D GS, designed for $\alpha$-blending approximation, necessitates pre-sorted Gaussians based on depth information derived from camera parameters. This poses a challenge for single images because detailed camera parameters are often not known in natural individual image, while non-natural images, including screenshots and AI-generated content, are not captured by cameras. Without accurate depth information, the Gaussian sorting might be impaired, diminishing the final fitting performance. 
Moreover, the current rasterization process skips the remaining Gaussians once the accumulated opacity surpasses the given threshold, which results in underutilization of Gaussian data, thereby requiring more Gaussians for high-quality rendering.

To address these issues, we propose a new paradigm of image representation and compression, namely GaussianImage, using 2D Gaussian Splatting. 
\textbf{Firstly}, we adopt 2D Gaussians in lieu of 3D for a compact and expressive representation. Each 2D Gaussian is defined by 4 attributes (9 parameters in total): position, anisotropic covariance, color coefficients, and opacity. This modification results in a 6.5$\times$ compression over 3D Gaussians with equivalent Gaussian points, significantly mitigating storage demands of Gaussian representation. 
\textbf{Subsequently}, we advocate a unique rasterization algorithm that replaces depth-based Gaussian sorting and $\alpha$-blending with a accumulated summation process. This novel approach directly computes each pixel's color from the weighted sum of 2D Gaussians, which not only fully utilizes the information of all Gaussian points covering the current pixel to improve fitting performance, but also avoids the tedious calculation of accumulated transparency to accelerate training and inference speed. More important, this summation mechanism allows us to merge color coefficients and opacity into a singular set of weighted color coefficients, reducing parameter count to 8 and further improving the compression ratio to 7.375$\times$. \textbf{Finally}, we transfer our 2D Gaussian representation into a practical image codec. Framing image compression as a Gaussian attribute compression task, we employ a two-step compression strategy: attribute quantization-aware fine-tuning and encoding. By applying 16-bit float quantization, 6-bit integer quantization \cite{bhalgat2020lsq+}, and residual vector quantization (RVQ) \cite{zeghidour2021soundstream} to positions, covariance parameters, and weighted color coefficients, respectively, we successfully develop the first image codec based on 2D Gaussian Splatting. As a preliminary proof of concept, the partial bits-back coding \cite{townsend2019practical, ryder2022split} is optionally used to further improve the compression performance of our codec. Overall, our contributions are threefold:
\begin{itemize}
    \item[$\bullet$] We present a pioneering paradigm of image representation and compression by 2D Gaussian Splatting. With compact 2D Gaussian representation and a novel accumulated blending-based rasterization method, our approach achieves high representation performance with short training duration, minimal GPU memory overhead and remarkably, 2000 FPS rendering speed. 
    \item[$\bullet$] We develop a low-complexity neural image codec using vector quantization. Furthermore, a partial bits-back coding technique is optionally used to reduce the bitrate.
    \item[$\bullet$] Experimental results show that when compared with existing INR methods, our approach achieves a remarkable training and inference acceleration with less GPU memory usage while maintaining similar visual quality. When used as an efficient image codec, our approach offers competitive compression performance comparable to COIN \cite{dupont2021coin} and COIN++ \cite{dupont2022coin++}. Comprehensive ablations and analyses demonstrate the effectiveness of each proposed component.
\end{itemize}

\section{Related Works}
\subsection{Implicit Neural Representation}
Recently, implicit neural representation has gained increasing attention for its wide-ranging potential applications, such as 3D scene rendering \cite{mildenhall2020nerf, barron2021mip, xu2022point, barron2022mip}, image \cite{sitzmann2020implicit, saragadam2023wire, muller2022instant, chen2023neurbf} and video \cite{chen2021nerv, li2022nerv, chen2023hnerv, zhang2024boosting} representations. We roughly classified existing image INRs into two categories: (i) MLP-based INRs \cite{sitzmann2020implicit, fathony2020multiplicative, tancik2020fourier, ramasinghe2022beyond, saragadam2023wire} take position encoding of spatial coordinates as input of an MLP network to learn the RGB values of images, while they only rely on the neural network to encode all the image information, resulting in inefficient training and inference especially for high-resolution image.(ii) Feature grid-based INRs \cite{martel2021acorn, takikawa2021neural, muller2022instant, chen2023neurbf} adopt a large-scale multi-resolution grid, such as quadtree and hash table, to provide prior information for a compact MLP. This reduces the learning difficulty of MLP to a certain extent and accelerates the training process, making INRs more practical. Unfortunately, they still consume large GPU memory, which is difficult to accommodate on low-end devices.
Instead of following existing INR methods, we aim to propose a brand-new image representation paradigm based on 2D Gaussian Splatting, which enables us to enjoy swifter training, faster rendering, and less GPU resource consumption.

\subsection{Gaussian Splatting}
Gaussian Splatting \cite{kerbl20233d} has recently gained tremendous traction as a promising paradigm to 3D view synthesis. With explicit 3D Gaussian representations and differentiable tile-based rasterization, GS not only brings unprecedented control and editability but also facilitates high-quality and real-time rendering in 3D scene reconstruction. This versatility has opened up new avenues in various domains, including simultaneous localization and mapping (SLAM) \cite{yan2023gs, huang2023photo, keetha2023splatam}, dynamic scene modeling \cite{wu20234d, luiten2023dynamic, yang2023real}, AI-generated content \cite{chen2024gaussianeditor, chen2024gen3d, zielonka2023drivable}, and autonomous driving \cite{zhou2023drivinggaussian, yan2024street}. Despite its great success in 3D scenarios, the application of GS to single image representation remains unexplored. Our work pioneers the adaptation of GS for 2D image representation, leveraging the strengths of GS in highly parallelized workflow and real-time rendering to outperform INR-based methods in terms of training efficiency and decoding speed.



\subsection{Image Compression}
Traditional image compression techniques, such as JPEG \cite{wallace1991jpeg}, JPEG2000 \cite{skodras2001jpeg} and BPG \cite{bpg2014}, follow a transformation, quantization, and entropy coding procedure to achieve good compression efficiency with decent decompression speed. Recently, learning-based image compression methods based on variational auto-encoder (VAE) have re-imagined this pipeline, integrating complex nonlinear transformations \cite{balle2015density, cheng2020learned, he2022elic, liu2023learned} and advanced entropy models \cite{balle2017end, balle2018variational, minnen2018joint, koyuncu2022contextformer}. Despite these methods surpassing traditional codecs in rate-distortion (RD) performance, their extremely high computational complexity and very slow decoding speed severely limit their practical deployment. 
To tackle the computational inefficiency of existing art, some works have explored INR-based compression methods \cite{dupont2021coin, dupont2022coin++, strumpler2022implicit, ladune2023cool, leguay2023low, guo2024compression}. However, as image resolutions climb, their decoding speeds falter dramatically, challenging their real-world applicability. In this paper, our approach diverges from VAE and INR paradigms, utilizing 2D Gaussian Splatting to forge a neural image codec with unprecedented decoding efficiency. This marks an important milestone for neural image codecs.

\section{Method}
\begin{figure}[t]
    \centering
    \includegraphics[width=1\textwidth]{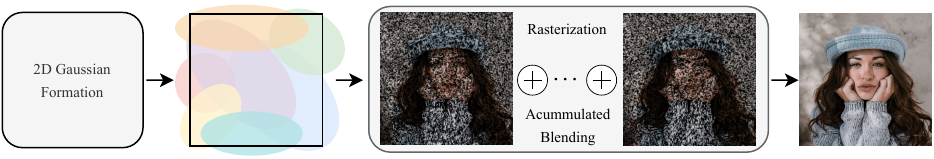}
    \caption{Our proposed GaussianImage framework. 2D Gaussians are first formatted and then rasterized to generate the output image. The rasterizer uses our proposed accumulated blending for efficient 2D image representation.}
    \label{fig:2DGS}
    \vspace{-0.5cm}
\end{figure}

Fig.~\ref{fig:2DGS} delineates the overall processing pipeline of our GaussianImage. Our approach begins by forming 2D Gaussians on an image plane, a process mainly calculating the 2D covariance matrix $\boldsymbol{\Sigma}$. Afterwards, we employ an accumulated blending mechanism to compute the value for each pixel. 
In what follows, we begin with the formation of a 2D Gaussian in Section~\ref{subsec:2d_gs_formation}. Next, we describe how to adapt the rasterization process of 3D GS to align the unique characteristics of 2D image representation and upgrade 2D Gaussian with less parameters in Section~\ref{subsec:rasterization}. Then, we present a two-step compression pipeline to convert our GaussianImage into a neural image codec in Section~\ref{subsec:compression_pipeline}. Finally, we state the training process on the image representation and compression tasks in Section~\ref{subsec:training}.

\subsection{2D Gaussian Formation}
\label{subsec:2d_gs_formation}
In 3D Gaussian Splatting, each 3D Gaussian is initially mapped into a 2D plane through viewing and projection transformation. Then a differentiable rasterizer is used to render the current view image from these projected Gaussians. Since our application is no longer oriented to 3D scenes, but to 2D image representation, we discard many bloated operations and redundant parameters in 3D GS, such as project transformation, spherical harmonics, etc. 

In our framework, the image representation unit is a 2D Gaussian. The basic 2D Gaussian is described by its position $\boldsymbol{\mu} \in \mathbb{R}^2$, 2D covariance matrix $\boldsymbol{\Sigma} \in \mathbb{R}^{2\times2}$, color coefficients $\boldsymbol{c} \in \mathbb{R}^{3}$ and opacity $o \in \mathbb{R}$. 
Note that covariance matrix $\boldsymbol{\Sigma}$ of a Gaussian distribution requires positive semi-definite. Typically, it is difficult to constrain the learnable parameters using gradient descent to generate such valid matrices. To avoid producing invalid matrix during training, we choose to optimize the factorized form of the covariance matrix. Here, we present two decomposition ways to cover all the information of the original covariance matrix. One intuitive decomposition is the Cholesky factorization \cite{higham2009cholesky}, which breaks down $\boldsymbol{\Sigma}$ into the product of a lower triangular matrix $\boldsymbol{L} \in \mathbb{R}^{2\times2}$ and its conjugate transpose $\boldsymbol{L}^T$:
\begin{equation}
  \begin{aligned}
    \boldsymbol{\Sigma} = \boldsymbol{L} \boldsymbol{L}^T.
    \label{eq:cholesky}
  \end{aligned}
\end{equation}
For the sake of writing, we use a Choleksy vector $\boldsymbol{l}=\{l_1, l_2, l_3\}$ to represent the lower triangular elements in matrix $\boldsymbol{L}$. When compared with 3D Gaussian having 59 learnable parameters, our 2D Gaussian only require 9 parameters, making it more lightweight and suitable for image representation.

Another decomposition follows 3D GS \cite{kerbl20233d} to factorize the covariance matrix into a rotation matrix $\boldsymbol{R} \in \mathbb{R}^{2\times2}$ and scaling matrix $\boldsymbol{S}\in \mathbb{R}^{2\times2}$:
\begin{equation}
  \begin{aligned}
    \boldsymbol{\Sigma} = (\boldsymbol{RS})(\boldsymbol{RS})^T,
    \label{eq:RSSR}
  \end{aligned}
\end{equation}
where the rotation matrix $\boldsymbol{R}$ and the scaling matrix $\boldsymbol{S}$ are expressed as
\begin{equation}
  \begin{aligned}
    \boldsymbol{R} = 
    \begin{bmatrix}
    \cos(\theta) & -\sin(\theta) \\
    \sin(\theta) & \cos(\theta)
    \end{bmatrix}, \quad
    \boldsymbol{S} = 
    \begin{bmatrix}
    s_{1} & 0 \\
    0 & s_{2}
    \end{bmatrix}.
    \label{eq:rot_scale}
  \end{aligned}
\end{equation}
Here, $\theta$ represents the rotation angle. $s_1$ and $s_2$ are scaling factors in different eigenvector directions. While the decomposition of the covariance matrix is not unique, they have equivalent capabilities to represent the image. However, the robustness to compression of different decomposition forms is inconsistent, which is explained in detail in the appendix. Therefore, we need to carefully choose the decomposition form of the covariance matrix when facing different image tasks.

\subsection{Accumulated Blending-based Rasterization}
\label{subsec:rasterization}
During the rasterization phase, 3D GS first forms a sorted list of Gaussians $\mathcal{N}$ based on the projected depth information. Then the $\alpha$-blending is adopted to render pixel $i$:
\begin{equation}
  \begin{aligned}
    \boldsymbol{C}_i = \sum_{n \in \mathcal{N}} \boldsymbol{c}_n \cdot \alpha_n \cdot T_n, \quad T_n = \prod_{m=1}^{n-1} (1 - \alpha_m),
    \label{eq:rasterization1}
  \end{aligned}
\end{equation}
where $T_n$ denotes the accumulated transparency. The $\alpha_n$ is computed with projected 2D covariance $\boldsymbol{\Sigma}$ and opacity $o_n$:
\begin{equation}
  \begin{aligned}
    \alpha_n = o_n \cdot \exp(-\sigma_n), \quad \sigma_n = \frac{1}{2} \boldsymbol{d}_n^T \boldsymbol{\Sigma}^{-1} \boldsymbol{d}_n,
    \label{eq:rasterization2}
  \end{aligned}
\end{equation} 
where $\boldsymbol{d} \in \mathbb{R}^2$ is the displacement between the pixel center and the projected 2D Gaussian center. 

Since the acquisition of depth information involves viewing transformation, it requires us to know the intrinsic and extrinsic parameters of the camera in advance. However, it is difficult for natural individual image to access the detailed camera parameters, while non-natural images, such as screenshots and AI-generated content, are not captured by the camera. In this case, retaining the $\alpha$-blending of the 3D GS without depth cues would result in arbitrary blending sequences, compromising the rendering quality. Moreover, 3D GS only maintains Gaussians with a 99\% confidence interval in order to solve the problem of numerical instability in computing the projected 2D covariance, but this makes only part of Gaussians covering pixel $i$ contribute to the rendering of pixel $i$, leading to inferior fitting performance. 

To overcome these limitations, we propose an accumulated summation mechanism to unleash the potential of our 2D Gaussian representation. Since there is no viewpoint influence when rendering an image, the rays we observe from each element are determined, and so as all the $\alpha$ values. Therefore, we merge the $T_n$ part in Equation~\ref{eq:rasterization1} into the $o_n$ term, and simplify the computation consuming $\alpha$-blending to a weighted sum:
\begin{equation}
  \begin{aligned}
    \boldsymbol{C}_i = \sum_{n \in \mathcal{N}} \boldsymbol{c}_n  \cdot \alpha_n = \sum_{n \in \mathcal{N}} \boldsymbol{c}_n  \cdot o_n \cdot \exp(-\sigma_n).
    \label{eq:rasterization3}
  \end{aligned}
\end{equation}
This removes the necessity of Gaussian sequence order, so that we can remove the sorting from rasterization.

\begin{figure}[t]
    \centering
    \includegraphics[width=0.9\textwidth]{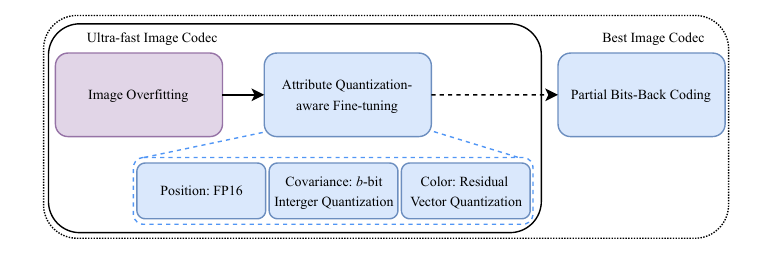}
    \vspace{-0.2cm}
    \caption{Compression pipeline of our proposed GaussianImage. After overfitting image, we apply attribute quantization-aware fine-tuning to build an ultra-fast image codec. Partial bits-back coding is used to achieve the best compression performance.}
    \label{fig:codec}
    \vspace{-0.2cm}
\end{figure}

This novel rasterization algorithm brings multiple benefits. First, our accumulated blending process is insensitive to the order of Gaussian points. This property allows us to avoid the impact of the random order of Gaussian points on rendering, achieving robustness to any order of Gaussian points. Second, when compared with Equation \ref{eq:rasterization1}, our rendering skips the tedious sequential calculation of accumulated transparency $T_n$, improving our training efficiency and rendering speed. Third, since the color coefficients $\boldsymbol{c}_n$ and the opacity $o_n$ are learnable parameters, they can be merged to further simplify Equation \ref{eq:rasterization3}: 
\begin{equation}
  \begin{aligned}
    \boldsymbol{C}_i = \sum_{n \in \mathcal{N}} \boldsymbol{c}_n' \cdot \exp(-\sigma_n),
    \label{eq:rasterization4}
  \end{aligned}
\end{equation}
where the weighted color coefficients $\boldsymbol{c}_n' \in \mathbb{R}^3$ is no longer limited in the range of $[0, 1]$. In this way, instead of the basic 2D Gaussian that requires 4 attributes in Section~\ref{subsec:2d_gs_formation}, our upgraded 2D Gaussian is described by only 3 attributes (i.e., position, covariance, and weighted color coefficients) with a total of 8 parameters. This further improves the compression ratio to 7.375$\times$ when compared with 3D Gaussian under equivalent Gaussian points.

\subsection{Compression Pipeline}
\label{subsec:compression_pipeline}
After overfitting the image, we propose a compression pipeline for image compression with GaussianImage. As shown in Fig.~\ref{fig:codec}, our standard compression pipeline is composed of two steps: image overfitting and attribute quantization-aware fine-tuning. To achieve the best compression performance, partial bits-back coding \cite{townsend2019practical, ryder2022split} is an optional strategy. Herein, we elucidate the compression process using our GaussianImage based on Cholesky factorization as an example.


\noindent \textbf{Attribute Quantization-aware Fine-tuning.} Given a set of 2D Gaussian points fit on an image, we apply distinct quantization strategies to various attributes. Since the Gaussian location is sensitive to quantization, we adopt 16-bit float precision for position parameters to preserve reconstruction fidelity. 
For Choleksy vector $\boldsymbol{l}_n$ in the $n$-th Gaussian, we incorporate a $b$-bit asymmetric quantization technique \cite{bhalgat2020lsq+}, where both the scaling factor $\gamma_{i}$ and the offset factor $\beta_{i}$ are learned during fine-tuning:
\begin{equation}
  \begin{aligned}
    \hat{l}_{i}^{n}= \left\lfloor \text{clamp}\left(\frac{l_{i}^{n}-\beta_{i}}{\gamma_{i}}, 0, 2^{b}-1\right) \right\rceil, \quad  \bar{l}_{i}^{n}= \hat{l}_{i}^{n} \times \gamma_{i} + \beta_{i},
  \end{aligned}
\end{equation} 
where $i \in \{0, 1, 2\}$. Note that we share the same scaling and offset factors at all Gaussians in order to reduce metadata overhead. After fine-tuning, the covariance parameters are encoded with $b$-bit precision, while the scaling and offset values required for re-scaling are stored in 32-bit float precision.  

As for weighted color coefficients, a codebook enables representative color attribute encoding via vector quantization (VQ) \cite{gray1984vector}. While naively applying vector quantization leads to inferior rendering quality, we employ residual vector quantization (RVQ) \cite{zeghidour2021soundstream} that cascades $M$ stages of VQ with codebook size $B$ to mitigate performance degradation:
\begin{equation}
  \begin{aligned}
    &\hat{\boldsymbol{c}}_n'^m= \sum_{k=1}^{m} \mathcal{C}^{k}[i^k], \quad m \in \{1, \cdots, M\}, \\
    &i_n^{m} = \mathop{\arg\min}\limits_{m} \left \| \mathcal{C}^{m}[k] - (\boldsymbol{c}_n' - \hat{\boldsymbol{c}}_n'^{m-1})  \right \|_2^2, \quad  \hat{\boldsymbol{c}}_n'^{0} = 0,
  \end{aligned}
\end{equation}
where $\hat{\boldsymbol{c}}_n'^m$ denotes the output color vector after $m$ quantization stages, $\mathcal{C}^m \in \mathbb{R}^{B\times3}$ represents the codebook at the stage $m$, $i^m \in \{0, \cdots, B-1\}^N$ is the codebook indices at the stage $m$, and $\mathcal{C}[i] \in \mathbb{R}^3$ is the vector at index $i$ of the codebook $\mathcal{C}$. To train the codebooks, we apply the commitment loss $\mathcal{L}_c$ as follows: 
\begin{equation}
  \begin{aligned}
    \mathcal{L}_c=\frac{1}{N\times B}\sum_{k=1}^{M}\sum_{n=1}^{N} \left \| {\rm sg}[c_n' - \hat{c}_n'^{k-1}] - \mathcal{C}^{k}[i_n^k] \right \|_2^2, 
  \end{aligned}
\end{equation}
where $N$ is the number of Gaussians and ${\rm sg}[\cdot]$ is the stop-gradient operation. 


\noindent \textbf{Partial Bits-Back Coding.} As we have not adopted any auto-regressive context~\cite{minnen2018joint} to encode 2D Gaussian parameters, any permutation of 2D Gaussian points can be seen as an equivariant graph without edge. Therefore, we can adopt bits-back coding \cite{townsend2019practical} for equivariant graph described by \cite{kunze2024entropy} to save bitrate. More specifically, \cite{kunze2024entropy} show that an unordered set with $N$ elements has $N!$ equivariant, and bits-back coding can save a bitrate of
\begin{gather}
    \log N! - \log N,
\end{gather}
compared with directly store those unordered elements. 

However, the vanilla bits-back coding requires initial bits \cite{townsend2019practical} of $\log N!$, which means that it can only work on a dataset, not on a single image. To tackle this challenge, \cite{ryder2022split} introduces a partial bits-back coding strategy that segments the image data, applying vanilla entropy coding to a fraction of the image as the initial bit allocation, with the remainder encoded via bits-back coding.


In our case, we reuse the idea of \cite{ryder2022split}. Specifically, we encode the initial $K$ Gaussians by vanilla entropy coding, and the subsequent $N-K$ Gaussians by bits-back coding. This segmented approach is applicable to single image compression, contingent upon the bitrate of the initial $K$ Gaussian exceeding the initial bits $\log (N-K)!$. Let $R_k$ denotes the bitrate of $k$-th Gaussian, the final bitrate saving can be formalized as:
\begin{gather}
    \log (N-K^*)! - \log (N-K^*), \\ \textrm{where } K^* = \inf K, \textrm{s.t.} \sum_{k=1}^K R_k - \log (N-K^*)! \ge 0.
\end{gather}
Despite its theoretical efficacy, bits-back coding may not align with the objective of developing an ultra-fast codec due to its slow processing latency \cite{kunze2024entropy}. Consequently, we leave this part as a preliminary proof of concept on the best rate-distortion performance our codec can achieve, instead of a final result of our codec can achieve with 2000 FPS.

\subsection{Training}
\label{subsec:training}
For image representation, our objective is to minimize the distortion between the original image $x$ and reconstructed image $\hat{x}$. To this end, we employ the L2 loss function to optimize the Gaussian parameters.
It is worth noting that previous GS method \cite{kerbl20233d} introduces adaptive density control to split and clone Gaussians when optimizing 3D scenes. Since there exists many empty areas in the 3D space, they need to consider avoiding populating these areas. By contrast, there is no so-called empty area in the 2D image space. Therefore, we discard adaptive density control, which greatly simplifies the optimization process of 2D image representation. 

As for image compression task, the overall loss $\mathcal{L}$ consists of the reconstruction loss $\mathcal{L}_{rec}$ and the commitment loss $\mathcal{L}_c$: 
\begin{equation}
  \begin{aligned}
    \mathcal{L}=\mathcal{L}_{rec}+ \lambda \mathcal{L}_c,
  \end{aligned}
\end{equation}
where $\lambda$ serves as a hyper-parameter, balancing the weight of each loss component. The color codebooks are initialized using the K-means algorithm, providing a robust starting point for subsequent optimization. During fine-tuning, we adopt the exponential moving average mode to update the codebook.



 \begin{table*}[t]
 \scriptsize
  \caption{Quantitative comparison with various baselines in PSNR, MS-SSIM, training time, rendering speed, GPU memory usage and parameter size.}
  \label{table:image_representation}
  \vspace{-0.5cm}
    \centering
    \begin{subtable}[t]{1\linewidth}
    \centering
    \caption{Kodak dataset}
     \vspace{-0.2cm}
        \begin{tabular}{l|cccccc}
            \toprule
            Methods & PSNR$\uparrow$ \quad & MS-SSIM$\uparrow$ \quad & Training Time(s)$\downarrow$ \quad & FPS$\uparrow$ \quad & GPU Mem(MiB)$\downarrow$ \quad &  Params(K)$\downarrow$ \\
            \midrule
            WIRE \cite{saragadam2023wire}   & 41.47 & 0.9939 & 14338.78 & 11.14 & 2619 &  136.74     \\
            SIREN \cite{sitzmann2020implicit}& 40.83 & 0.9960 & 6582.36 & 29.15 & 1809 & 272.70      \\
            I-NGP \cite{muller2022instant}  & \cellcolor{lightyellow}43.88 & 0.9976 & 490.61 & \cellcolor{lightyellow}1296.82 & 1525 &300.09 \\
            NeuRBF \cite{chen2023neurbf}    & 43.78 &  0.9964 & 991.83 & 663.01 & 2091 & 337.29      \\
            3D GS \cite{kerbl20233d}        & 43.69 & \cellcolor{lightred}0.9991 & \cellcolor{lightyellow}339.78 & 859.44   & \cellcolor{lightyellow}557 &  3540.00 \\ 
            Ours                            & \cellcolor{lightred}44.08 & \cellcolor{lightyellow}0.9985 & \cellcolor{lightred}106.59 & \cellcolor{lightred}2092.17   & \cellcolor{lightred}419 & 560.00  \\
            \bottomrule
        \end{tabular}
    \end{subtable}\\
    \begin{subtable}[t]{1\linewidth}
    \centering
    \caption{DIV2K dataset}
    \vspace{-0.2cm}
        \begin{tabular}{l|cccccc}
            \toprule
            Methods & PSNR$\uparrow$ \quad & MS-SSIM$\uparrow$ \quad & Training Time(s)$\downarrow$ \quad & FPS$\uparrow$ \quad & GPU Mem(MiB)$\downarrow$ \quad &  Params(K)$\downarrow$ \\
            \midrule
            WIRE \cite{saragadam2023wire}   & 35.64 & 0.9511 & 25684.23 & 14.25 & 2619 & 136.74     \\
            SIREN \cite{sitzmann2020implicit} & 39.08 & 0.9958 & 15125.11 & 11.07 & 2053 & 483.60     \\
            I-NGP \cite{muller2022instant}  & 37.06 & 0.9950 & 676.29 & \cellcolor{lightyellow}1331.54 & 1906 & 525.40  \\
            NeuRBF \cite{chen2023neurbf}    & 38.60 & 0.9913 & 1715.44 & 706.40 & 2893 &383.65       \\
            3D GS \cite{kerbl20233d}        & \cellcolor{lightyellow}39.36 & \cellcolor{lightred}0.9979 & \cellcolor{lightyellow}481.27 & 640.33  & \cellcolor{lightyellow}709 & 4130.00     \\
            Ours                            & \cellcolor{lightred}39.53 & \cellcolor{lightyellow}0.9975 & \cellcolor{lightred}120.76 & \cellcolor{lightred}1737.60 & \cellcolor{lightred}439  & 560.00      \\
            \bottomrule
        \end{tabular}
    \end{subtable}
    \label{tab:array}
    \vspace{-0.3cm}
\end{table*}

\begin{figure*}[t]
  \centering
  \subfloat
  {\includegraphics[scale=0.4]{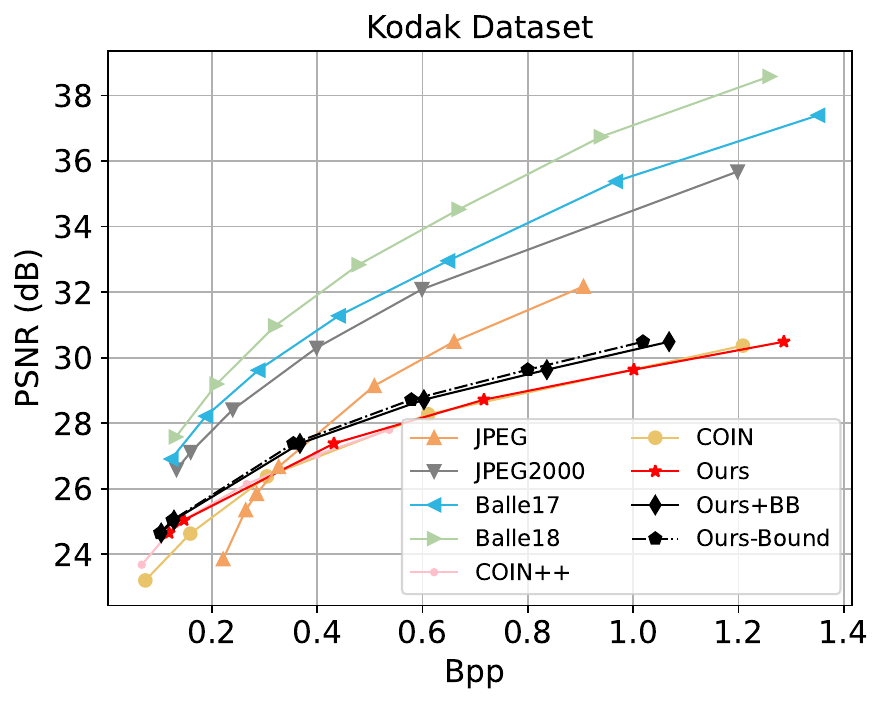}}
  \subfloat
  {\includegraphics[scale=0.4]{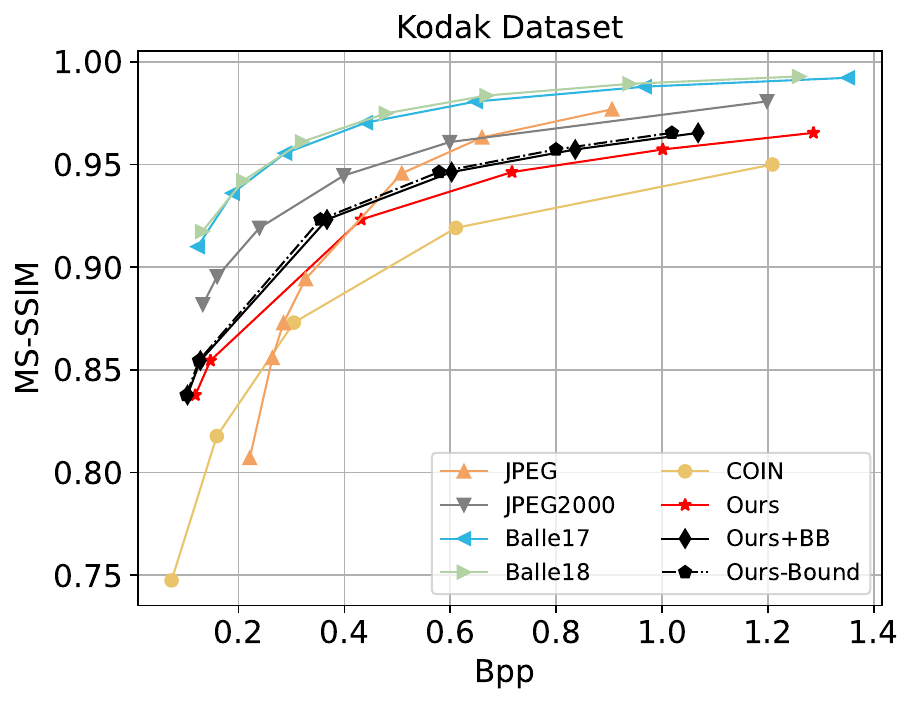}}\\
  \subfloat
  {\includegraphics[scale=0.4]{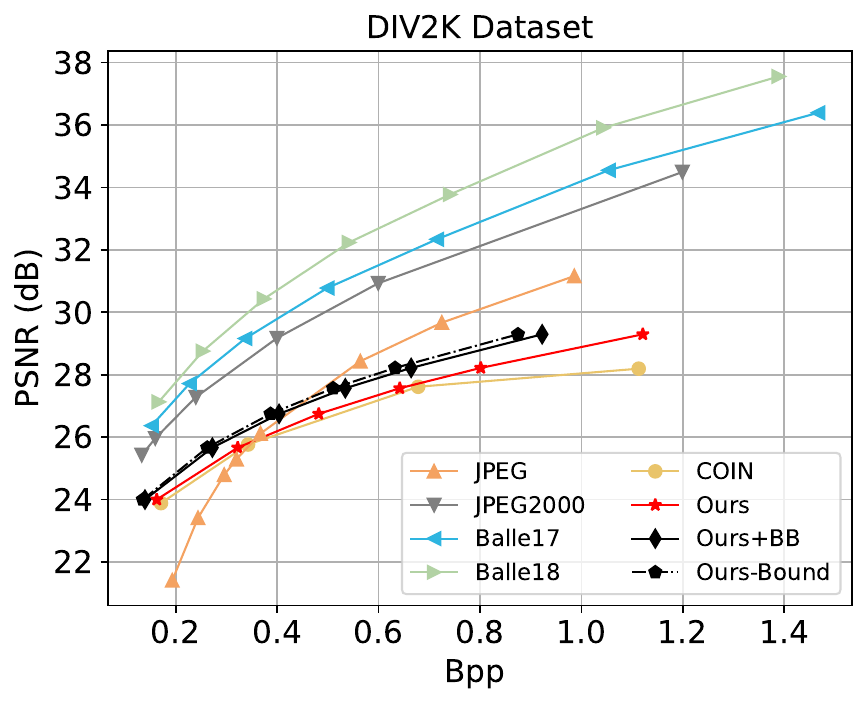}}
  \subfloat
  {\includegraphics[scale=0.4]{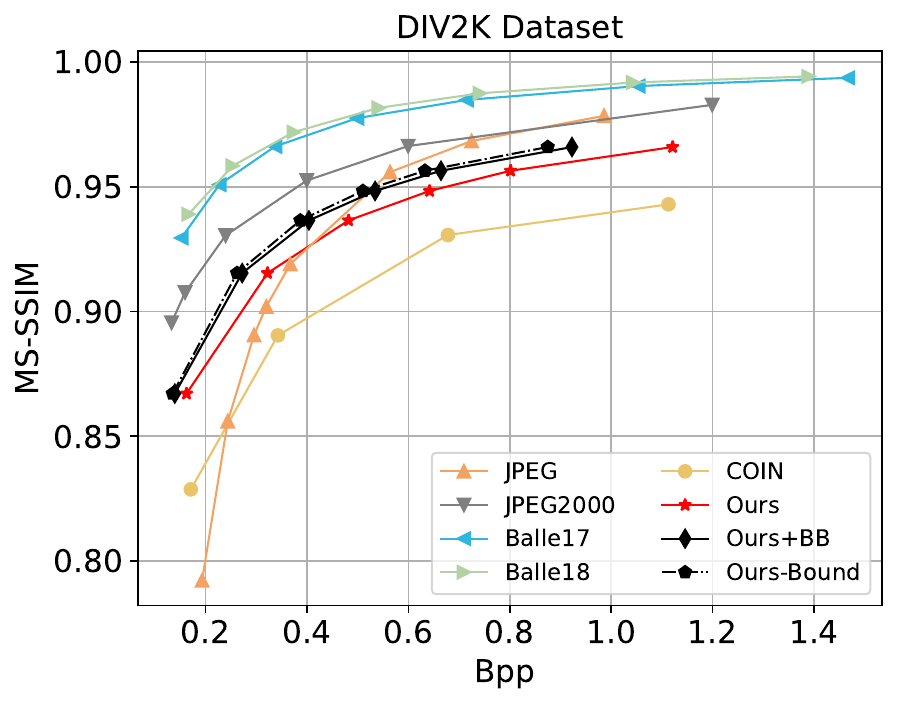}}
  \vspace{-0.2cm}
  \caption{Rate-distortion curves of our approach and different baselines on Kodak and DIV2K datasets in PSNR and MS-SSIM. BB denotes partial bits-back coding. Bound denotes the theoretical rate of our codec.}
  \label{fig:rd_performance}
  \vspace{-0.4cm}
\end{figure*}

\section{Experiments}
\subsection{Experimental Setup}

\textbf{Dataset.} Our evaluation in image representation and compression is conducted on two popular datasets. We use the Kodak dataset \cite{kodak}, which consists of 24 images with a resolution of 768$\times$512, and the DIV2K validation set \cite{Agustsson2017} with 2$\times$ bicubic downscaling, featuring 100 images with dimensions varying from 408$\times$1020 to 1020$\times$1020.


\noindent \textbf{Evaluation Metrics.} To assess image quality, we employ two esteemed metrics: PSNR and MS-SSIM \cite{wang2003multiscale}, which measure the distortion between reconstructed images and their originals. The bitrate for image compression is quantified in bits per pixel (bpp).


\noindent \textbf{Implementation Details.} Our GaussianImage, developed on top of gsplat \cite{gsplat2023}, incorporates custom CUDA kernels for rasterization based on accumulated blending. We represent the covariance of 2D Gaussians using Cholesky factorization unless otherwise stated. The Gaussian parameters are optimized over 50000 steps using the Adan optimizer \cite{xie2022adan}, starting with an initial learning rate of $1e^{-3}$, halved every 20000 steps. During attribute quantization-aware fine-tuning, the quantization precision $b$ of covariance parameters is set to 6 bits, with the RVQ color vectors' codebook size $B$ and the number of quantization stages $M$ fixed at 8 and 2, respectively. The iterations of K-means algorithm are set to 5. Experiments are performed using NVIDIA V100 GPUs and PyTorch, with further details available in the supplementary material.


\noindent \textbf{Benchmarks.} For image representation comparisons, GaussianImage is benchmarked against competitive INR methods like SIREN \cite{sitzmann2020implicit}, WIRE \cite{saragadam2023wire}, I-NGP \cite{muller2022instant}, and NeuRBF \cite{chen2023neurbf}. As for image compression, baselines span traditional codecs (JPEG \cite{wallace1991jpeg}, JPEG2000 \cite{skodras2001jpeg}), VAE-based codecs (Ball{\'e}17 \cite{balle2017end}, Ball{\'e}18 \cite{balle2018variational}), INR-based codecs (COIN \cite{dupont2021coin}, COIN$++$ \cite{dupont2022coin++}). We utilize the open-source PyTorch implementation \cite{neurbf2023} of NeuRBF for I-NGP. These INR methods maintain consistent training steps with GaussianImage. Detailed implementation notes for baselines are found in the appendix.


\subsection{Image Representation}
Fig.~\ref{fig:image_representation} (left) and \tableautorefname~\ref{table:image_representation} show the representation performance of various methods on the Kodak and DIV2K datasets under the same training steps. Although MLP-based INR methods (SIREN \cite{sitzmann2020implicit}, WIRE \cite{saragadam2023wire}) utilize fewer parameters to fit an image, they suffer from enormous training time and hyperslow rendering speed. Recent feature grid-based INR methods (I-NGP \cite{muller2022instant}, NeuRBF \cite{chen2023neurbf}) accelerate training and inference, but they demand substantially more GPU memory compared to GS-based methods.
Since the original 3D GS uses 3D Gaussian as the representation unit, it face the challenge of giant parameter count, which decelerates training and restricts inference speed. By choosing 2D Gaussian as the representation unit, our method secures pronounced advantages in training duration, rendering velocity, and GPU memory usage, while substantially reducing the number of stored parameters yet preserving comparable fitting quality.

\begin{figure}[t]
    \centering
    \includegraphics[width=1\textwidth]{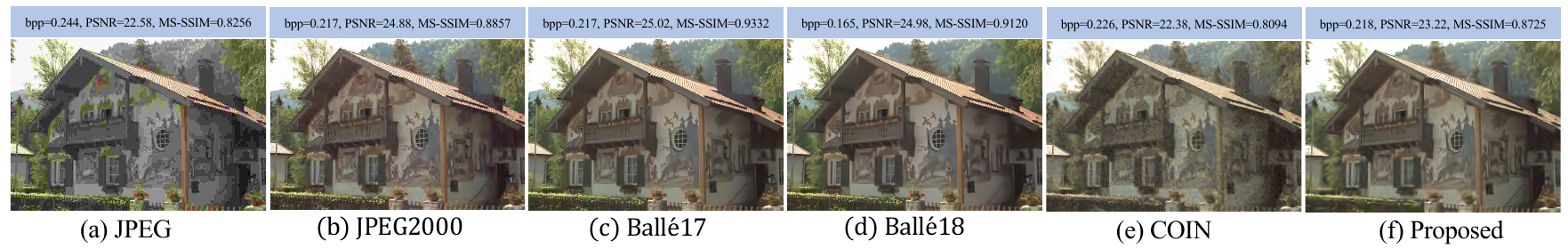}
    \vspace{-0.4cm}
    \caption{Subjective comparison of various codecs on Kodak at low Bpp.}
    \label{fig:visualization}
    \vspace{-0.2cm}
\end{figure}

\begin{table*}[t]
 \scriptsize
  \caption{Computational complexity of traditional and learning-based image codecs on DIV2K Dataset at low and high Bpp.}
  \label{table:computational_complexity}
  \centering
  \vspace{-0.2cm}
  \begin{tabular}{l|c|c|c|c|c}
    \toprule
    Methods & Bpp$\downarrow$ \quad & PSNR$\uparrow$ \quad & MS-SSIM$\uparrow$ \quad & Encoding FPS$\uparrow$ \quad & Decoding FPS$\uparrow$ \quad \\
    \midrule
    JPEG  \cite{wallace1991jpeg}      & 0.3197/0.5638 & 25.2920/28.4299  & 0.9020/0.9559  & \cellcolor{lightred}608.61/557.35 & \cellcolor{lightyellow}614.68/545.59 \\
    JPEG2000  \cite{skodras2001jpeg}  & 0.2394/0.5993 & 27.2792/30.9294 & 0.9305/0.9663  & 3.46/3.40 & 4.32/3.93 \\
    Ball{\'e}17 \cite{balle2017end}  & 0.2271/0.4987 & 27.7168/30.7759 & 0.9508/0.9775 & \cellcolor{lightyellow}21.23/16.53 & 18.83/17.87 \\
    Ball{\'e}18  \cite{balle2018variational} & 0.2533/0.5415 & 28.7548/32.2351 & 0.9584/0.9816  & 16.53/13.56 & 15.87/15.20 \\
    COIN  \cite{dupont2021coin}      & 0.3419/0.6780 & 25.8012/27.6126 & 0.8905/0.9306 & 5.30$e^{-4}$/3.51$e^{-4}$ & 166.31/93.74 \\
    Ours        & 0.3221/0.6417 & 25.6631/27.5656 & 0.9154/0.9483 & 4.11$e^{-3}$/4.73$e^{-3}$ & \cellcolor{lightred}1970.76/1980.54 \\ 
    \bottomrule
  \end{tabular}
  \vspace{-0.2cm}
\end{table*}

\subsection{Image Compression}
\noindent \textbf{Coding performance.}
Fig.~\ref{fig:rd_performance} presents the RD curves of various codecs on the Kodak and DIV2K datasets. 
Notably, our method achieves comparable compression performance with COIN \cite{dupont2021coin} and COIN++ \cite{dupont2022coin++} in PSNR. With the help of the partial bits-back coding, our codec can outperform COIN and COIN++. Furthermore, when measured by MS-SSIM, our method surpasses COIN by a large margin. 
Fig.~\ref{fig:visualization} provides a qualitative comparison between our method, JPEG \cite{wallace1991jpeg}, and COIN, revealing that our method restores image details more effectively and delivers superior reconstruction quality by consuming lower bits.




\noindent \textbf{Computational complexity.} \tableautorefname~\ref{table:computational_complexity} reports the computational complexity of several image codecs on the DIV2K dataset, with learning-based codecs operating on an NVIDIA V100 GPU and traditional codecs running on an Intel Core(TM) i9-10920X processor at a base frequency of 3.50GHz in single-thread mode. Impressively, the decoding speed of our codec reaches around 2000 FPS, outpacing traditional codecs like JPEG, while also providing enhanced compression performance at lower bitrates. This establishes a significant advancement in the field of neural image codecs.




\begin{table*}[t]
 \scriptsize
  \caption{Ablation study of image representation on Kodak dataset with 30000 Gaussian points over 50000 training steps. AR means accumulated blending-based rasterization, M indicates merging color coefficients $\boldsymbol{c}$ and opacity $o$. RS denotes decomposing the covariance matrix into rotation and scaling matrices. The final row in each subclass represents our default solution. }
  \label{table:ablation_representation}
  \centering
  \vspace{-0.2cm}
    \begin{tabular}{l|ccccc}
        \toprule
        Methods & PSNR$\uparrow$ \quad & MS-SSIM$\uparrow$ \quad & Training Time(s)$\downarrow$ \quad & FPS$\uparrow$ \quad & Params(K)$\downarrow$ \quad \\
        \midrule
        3D GS (w/ L1+SSIM)      & 37.75 & 0.9961 & 285.26 & 1067 & 1770      \\
        3D GS (w/ L2)           & 37.41 & 0.9947 & 197.90 & 1190 & 1770      \\
        Ours (w/ L2+w/o AR+w/o M)                & 37.89 & 0.9961 & 104.76 & 2340 & 270      \\
        Ours (w/ L2+w/ AR+w/o M)           & \cellcolor{lightred}38.69 & \cellcolor{lightred}0.9963 & \cellcolor{lightyellow}98.54 & \cellcolor{lightyellow}2555 & 270    \\
        Ours(w/ L2+w/ AR+w/ M) & \cellcolor{lightyellow}38.57 & \cellcolor{lightyellow}0.9961 & \cellcolor{lightred}91.06 & \cellcolor{lightred}2565 &  240     \\
        \midrule
        Ours (w/ L1)       & 36.46 & 0.9937 & \cellcolor{lightyellow}92.68 & 2438 &  240  \\
        Ours (w/ SSIM)       & 35.65 & \cellcolor{lightyellow}0.9952 & 183.20 & 2515 &  240     \\
        Ours (w/ L1+SSIM)       & \cellcolor{lightyellow}36.57 & 0.9945 & 188.22 & \cellcolor{lightred}2576 &  240     \\
        Ours (w/ L2+SSIM)       &  34.73 & 0.9932 &  189.17 & 2481 &  240     \\
        Ours (w/ L2) & \cellcolor{lightred}38.57 & \cellcolor{lightred}0.9961 & \cellcolor{lightred}91.06 & \cellcolor{lightyellow}2565 &  240     \\
        \midrule
        Ours-RS     & 38.83 & 0.9964 & 98.55 & 2321 &  240     \\
        Ours-Cholesky & 38.57 & 0.9961 & 91.06 & 2565 &  240     \\
        \bottomrule
    \end{tabular}
  \vspace{-0.2cm}
\end{table*}

\subsection{Ablation Study}
\noindent\textbf{Effect of different components.} 
To highlight the contributions of the key components in GaussianImage, we conduct a comprehensive set of ablation studies, as detailed in \tableautorefname~\ref{table:ablation_representation}. Initially, the original 3D GS \cite{kerbl20233d} method, which employs a combination of L1 and SSIM loss, is adapted to use L2 loss. This modification halves the training time at a minor cost to performance. Then, we replace the 3D Gaussian with the basic 2D Gaussian in Section \ref{subsec:2d_gs_formation}, which not only improves the fitting performance and decreases training time by $\frac{1}{3}$, but also doubles the inference speed and reduces parameter count by 6.5$\times$. By simplifying alpha blending to accumulated blending, we eliminate the effects of random 2D Gaussian ordering and bypasses the complex calculations for the accumulated transparency $T$, resulting in a significant 0.8dB improvement in PSNR alongside notable training and inference speed gains. This underscores the efficiency of our proposed accumulated blending approach. Furthermore, by merging the color vector $\boldsymbol{c}$ and opacity $o$ to form our upgraded 2D Gaussian, we observe a 10\% reduction in parameter count with a negligible 0.1dB decrease in PSNR.



\noindent \textbf{Loss function.} We evaluate various combinations of L2, L1, and SSIM losses, with findings presented in \tableautorefname~\ref{table:ablation_representation}. These results confirm that L2 loss is optimally suited for our approach, significantly improving image reconstruction quality while facilitating rapid training.



\noindent \textbf{Factorized form of covariance matrix.} As outlined in Section \ref{subsec:2d_gs_formation}, we optimize the factorized form of the covariance matrix through decomposition. The findings detailed in \tableautorefname~\ref{table:ablation_representation} demonstrate that various factorized forms possess similar capabilities in representing images, despite the decomposition's inherent non-uniqueness. The appendix provides additional analysis on the compression robustness of different factorized forms.

\noindent\textbf{Quantization strategies.} \tableautorefname~\ref{table:ablation_quantization} investigates the effect of different quantization schemes on image compression. Without the commitment loss $\mathcal{L}_c$ (V1), the absence of supervision for the RVQ codebook leads to significant deviations of the codebook vector from the original vector, adversely affecting reconstruction quality. Moreover, eliminating RVQ in favor of 6-bit integer quantization for color parameters (V2) resulted in a 7.02\% increase in bitrate consumption when compared with our default solution. This outcome suggests that the color vectors across different Gaussians share similarities, making them more suitable for RVQ. Further exploration into the use of higher bit quantization (V3) reveals a deterioration in compression efficiency.


\begin{table*}[t]
 \scriptsize
  \caption{Ablation study of quantization schemes on Kodak dataset. The first row denotes our final solution and is set as the anchor.}
  \label{table:ablation_quantization}
  \centering
  \vspace{-0.2cm}
  \begin{tabular}{l|cc|cc}
    \toprule
    Variants  & BD-PSNR (dB) $\uparrow$ & BD-rate (\%) $\downarrow$ & BD-MS-SSIM $\uparrow$ & BD-rate (\%) $\downarrow$ \\
    \midrule
    Ours                                        & 0 & 0 & 0 & 0 \\
    (V1) w/o $\mathcal{L}_c$+w/ RVQ + 6bit      & -3.145 & 333.16 & -0.0824 & 337.84 \\
    (V2) w/o $\mathcal{L}_c$+w/o RVQ + 6bit     & -0.159 & 7.02 & -0.0030 & 6.14 \\
    (V3) w/o $\mathcal{L}_c$+w/o RVQ + 8bit     & -0.195 & 11.69 & -0.0127 & 62.77 \\
    \bottomrule
  \end{tabular}
  \vspace{-0.2cm}
\end{table*}

\section{Conclusion}
In this work, we introduce GaussianImage, an innovative paradigm for image representation that leverages 2D Gaussian Splatting. This approach diverges significantly from the commonly utilized implicit neural networks, offering a discrete and explicit representation of images. When compared to 3D Gaussian Splatting, employing 2D Gaussian kernels brings forth two notable benefits for image representation. Firstly, the computationally intensive alpha blending is simplified to an efficient and permutation-invariant accumulated summation blending. Secondly, the quantity of parameters required for each Gaussian diminishes drastically from 59 to just 8, marking a substantial reduction in complexity. Consequently, GaussianImage emerges as a highly efficient and compact technique for image coding. Experimental results confirm that this explicit representation strategy enhances training and inference efficiency substantially. Moreover, it delivers a competitive rate-distortion performance after adopting vector quantization on parameters, compared to methods adopting implicit neural representation. These findings suggest promising avenues for further exploration in non-end-to-end image compression and representation strategies.

\noindent \textbf{Acknowledgments.} 
This work was supported by the National Natural Science Fund of China (Project No. 42201461, 62102024, 62331014) and the General Research Fund (Project No. 16209622) from the Hong Kong Research Grants Council.

\bibliographystyle{splncs04}
\bibliography{main}

\begin{thebibliography}{10}
\providecommand{\url}[1]{\texttt{#1}}
\providecommand{\urlprefix}{URL }
\providecommand{\doi}[1]{https://doi.org/#1}

\bibitem{kodak}
Kodak lossless true color image suite. \url{https://r0k.us/graphics/kodak/} (1999)

\bibitem{Agustsson2017}
Agustsson, E., Timofte, R.: Ntire 2017 challenge on single image super-resolution: Dataset and study. In: The IEEE Conference on Computer Vision and Pattern Recognition (CVPR) Workshops (July 2017)

\bibitem{ahmed1974discrete}
Ahmed, N., Natarajan, T., Rao, K.R.: Discrete cosine transform. IEEE transactions on Computers  \textbf{100}(1),  90--93 (1974)

\bibitem{balle2015density}
Ball{\'e}, J., Laparra, V., Simoncelli, E.P.: Density modeling of images using a generalized normalization transformation. arXiv preprint arXiv:1511.06281  (2015)

\bibitem{balle2017end}
Ball{\'e}, J., Laparra, V., Simoncelli, E.P.: End-to-end optimized image compression. In: International Conference on Learning Representations (2017)

\bibitem{balle2018variational}
Ball{\'e}, J., Minnen, D., Singh, S., Hwang, S.J., Johnston, N.: Variational image compression with a scale hyperprior. In: International Conference on Learning Representations (2018)

\bibitem{barron2021mip}
Barron, J.T., Mildenhall, B., Tancik, M., Hedman, P., Martin-Brualla, R., Srinivasan, P.P.: Mip-nerf: A multiscale representation for anti-aliasing neural radiance fields. In: Proceedings of the IEEE/CVF International Conference on Computer Vision. pp. 5855--5864 (2021)

\bibitem{barron2022mip}
Barron, J.T., Mildenhall, B., Verbin, D., Srinivasan, P.P., Hedman, P.: Mip-nerf 360: Unbounded anti-aliased neural radiance fields. In: Proceedings of the IEEE/CVF Conference on Computer Vision and Pattern Recognition. pp. 5470--5479 (2022)

\bibitem{begaint2020compressai}
B{\'e}gaint, J., Racap{\'e}, F., Feltman, S., Pushparaja, A.: Compressai: a pytorch library and evaluation platform for end-to-end compression research. arXiv preprint arXiv:2011.03029  (2020)

\bibitem{bpg2014}
Bellard, F.: Bpg image format. \url{https://bellard.org/bpg/} (2014)

\bibitem{bhalgat2020lsq+}
Bhalgat, Y., Lee, J., Nagel, M., Blankevoort, T., Kwak, N.: Lsq+: Improving low-bit quantization through learnable offsets and better initialization. In: Proceedings of the IEEE/CVF Conference on Computer Vision and Pattern Recognition Workshops. pp. 696--697 (2020)

\bibitem{chen2024survey}
Chen, G., Wang, W.: A survey on 3d gaussian splatting. arXiv preprint arXiv:2401.03890  (2024)

\bibitem{chen2023hnerv}
Chen, H., Gwilliam, M., Lim, S.N., Shrivastava, A.: Hnerv: A hybrid neural representation for videos. In: Proceedings of the IEEE/CVF Conference on Computer Vision and Pattern Recognition. pp. 10270--10279 (2023)

\bibitem{chen2021nerv}
Chen, H., He, B., Wang, H., Ren, Y., Lim, S.N., Shrivastava, A.: Nerv: Neural representations for videos. Advances in Neural Information Processing Systems  \textbf{34},  21557--21568 (2021)

\bibitem{chen2021learning}
Chen, Y., Liu, S., Wang, X.: Learning continuous image representation with local implicit image function. In: Proceedings of the IEEE/CVF conference on computer vision and pattern recognition. pp. 8628--8638 (2021)

\bibitem{chen2024gaussianeditor}
Chen, Y., Chen, Z., Zhang, C., Wang, F., Yang, X., Wang, Y., Cai, Z., Yang, L., Liu, H., Lin, G.: Gaussianeditor: Swift and controllable 3d editing with gaussian splatting. In: IEEE Conference on Computer Vision and Pattern Recognition (CVPR) (2024)

\bibitem{neurbf2023}
Chen, Z., Li, Z., Song, L., Chen, L., Yu, J., Yuan, J., Xu, Y.: \url{https://github.com/oppo-us-research/NeuRBF}

\bibitem{chen2023neurbf}
Chen, Z., Li, Z., Song, L., Chen, L., Yu, J., Yuan, J., Xu, Y.: Neurbf: A neural fields representation with adaptive radial basis functions. In: Proceedings of the IEEE/CVF International Conference on Computer Vision. pp. 4182--4194 (2023)

\bibitem{chen2024gen3d}
Chen, Z., Wang, F., Liu, H.: Text-to-3d using gaussian splatting. In: IEEE Conference on Computer Vision and Pattern Recognition (CVPR) (2024)

\bibitem{cheng2020learned}
Cheng, Z., Sun, H., Takeuchi, M., Katto, J.: Learned image compression with discretized gaussian mixture likelihoods and attention modules. In: Proceedings of the IEEE/CVF Conference on Computer Vision and Pattern Recognition. pp. 7939--7948 (2020)

\bibitem{Duda2013AsymmetricNS}
Duda, J.: Asymmetric numeral systems: entropy coding combining speed of huffman coding with compression rate of arithmetic coding. arXiv: Information Theory  (2013), \url{https://api.semanticscholar.org/CorpusID:13409455}

\bibitem{dupont2022coin++}
Dupont, E., Loya, H., Alizadeh, M., Golinski, A., Teh, Y., Doucet, A.: Coin++: neural compression across modalities. Transactions on Machine Learning Research  \textbf{2022}(11) (2022)

\bibitem{dupont2021coin}
Dupont, E., Golinski, A., Alizadeh, M., Teh, Y.W., Doucet, A.: Coin: Compression with implicit neural representations. In: Neural Compression: From Information Theory to Applications--Workshop@ ICLR 2021 (2021)

\bibitem{fang2023gaussianeditor}
Fang, J., Wang, J., Zhang, X., Xie, L., Tian, Q.: Gaussianeditor: Editing 3d gaussians delicately with text instructions. arXiv preprint arXiv:2311.16037  (2023)

\bibitem{fathony2020multiplicative}
Fathony, R., Sahu, A.K., Willmott, D., Kolter, J.Z.: Multiplicative filter networks. In: International Conference on Learning Representations (2020)

\bibitem{gray1984vector}
Gray, R.: Vector quantization. IEEE Assp Magazine  \textbf{1}(2),  4--29 (1984)

\bibitem{guo2024compression}
Guo, Z., Flamich, G., He, J., Chen, Z., Hern{\'a}ndez-Lobato, J.M.: Compression with bayesian implicit neural representations. Advances in Neural Information Processing Systems  \textbf{36} (2024)

\bibitem{he2022elic}
He, D., Yang, Z., Peng, W., Ma, R., Qin, H., Wang, Y.: Elic: Efficient learned image compression with unevenly grouped space-channel contextual adaptive coding. In: Proceedings of the IEEE/CVF Conference on Computer Vision and Pattern Recognition. pp. 5718--5727 (2022)

\bibitem{he2022poelic}
He, D., Yang, Z., Yu, H., Xu, T., Luo, J., Chen, Y., Gao, C., Shi, X., Qin, H., Wang, Y.: Po-elic: Perception-oriented efficient learned image coding. In: Proceedings of the IEEE/CVF Conference on Computer Vision and Pattern Recognition. pp. 1764--1769 (2022)

\bibitem{heil1989continuous}
Heil, C.E., Walnut, D.F.: Continuous and discrete wavelet transforms. SIAM review  \textbf{31}(4),  628--666 (1989)

\bibitem{higham2009cholesky}
Higham, N.J.: Cholesky factorization. Wiley interdisciplinary reviews: computational statistics  \textbf{1}(2),  251--254 (2009)

\bibitem{hu2020towards-icm}
Hu, Y., Yang, S., Yang, W., Duan, L.Y., Liu, J.: Towards coding for human and machine vision: A scalable image coding approach. In: 2020 IEEE International Conference on Multimedia and Expo (ICME). pp.~1--6. IEEE (2020)

\bibitem{huang2023photo}
Huang, H., Li, L., Cheng, H., Yeung, S.K.: Photo-slam: Real-time simultaneous localization and photorealistic mapping for monocular, stereo, and rgb-d cameras. arXiv preprint arXiv:2311.16728  (2023)

\bibitem{keetha2023splatam}
Keetha, N., Karhade, J., Jatavallabhula, K.M., Yang, G., Scherer, S., Ramanan, D., Luiten, J.: Splatam: Splat, track \& map 3d gaussians for dense rgb-d slam. arXiv preprint arXiv:2312.02126  (2023)

\bibitem{kerbl20233d}
Kerbl, B., Kopanas, G., Leimk{\"u}hler, T., Drettakis, G.: 3d gaussian splatting for real-time radiance field rendering. ACM Transactions on Graphics  \textbf{42}(4) (2023)

\bibitem{koyuncu2022contextformer}
Koyuncu, A.B., Gao, H., Boev, A., Gaikov, G., Alshina, E., Steinbach, E.: Contextformer: A transformer with spatio-channel attention for context modeling in learned image compression. In: European Conference on Computer Vision. pp. 447--463. Springer (2022)

\bibitem{kunze2024entropy}
Kunze, J., Severo, D., Zani, G., van~de Meent, J.W., Townsend, J.: Entropy coding of unordered data structures. In: The Twelfth International Conference on Learning Representations (2024), \url{https://openreview.net/forum?id=afQuNt3Ruh}

\bibitem{ladune2023cool}
Ladune, T., Philippe, P., Henry, F., Clare, G., Leguay, T.: Cool-chic: Coordinate-based low complexity hierarchical image codec. In: Proceedings of the IEEE/CVF International Conference on Computer Vision. pp. 13515--13522 (2023)

\bibitem{leguay2023low}
Leguay, T., Ladune, T., Philippe, P., Clare, G., Henry, F., D{\'e}forges, O.: Low-complexity overfitted neural image codec. In: 2023 IEEE 25th International Workshop on Multimedia Signal Processing (MMSP). pp.~1--6. IEEE (2023)

\bibitem{li2022nerv}
Li, Z., Wang, M., Pi, H., Xu, K., Mei, J., Liu, Y.: E-nerv: Expedite neural video representation with disentangled spatial-temporal context. In: European Conference on Computer Vision. pp. 267--284. Springer (2022)

\bibitem{liu2023learned}
Liu, J., Sun, H., Katto, J.: Learned image compression with mixed transformer-cnn architectures. In: Proceedings of the IEEE/CVF Conference on Computer Vision and Pattern Recognition. pp. 14388--14397 (2023)

\bibitem{luiten2023dynamic}
Luiten, J., Kopanas, G., Leibe, B., Ramanan, D.: Dynamic 3d gaussians: Tracking by persistent dynamic view synthesis. arXiv preprint arXiv:2308.09713  (2023)

\bibitem{ma2022recovering}
Ma, C., Yu, P., Lu, J., Zhou, J.: Recovering realistic details for magnification-arbitrary image super-resolution. IEEE Transactions on Image Processing  \textbf{31},  3669--3683 (2022)

\bibitem{martel2021acorn}
Martel, J.N., Lindell, D.B., Lin, C.Z., Chan, E.R., Monteiro, M., Wetzstein, G.: Acorn: adaptive coordinate networks for neural scene representation. ACM Transactions on Graphics (TOG)  \textbf{40}(4),  1--13 (2021)

\bibitem{mentzer2020hific}
Mentzer, F., Toderici, G.D., Tschannen, M., Agustsson, E.: High-fidelity generative image compression. Advances in Neural Information Processing Systems  \textbf{33},  11913--11924 (2020)

\bibitem{mildenhall2020nerf}
Mildenhall, B., Srinivasan, P.P., Tancik, M., Barron, J.T., Ramamoorthi, R., Ng, R.: Nerf: Representing scenes as neural radiance fields for view synthesis. In: European Conference on Computer Vision. pp. 405--421. Springer (2020)

\bibitem{minnen2018joint}
Minnen, D., Ball{\'e}, J., Toderici, G.D.: Joint autoregressive and hierarchical priors for learned image compression. In: Advances in neural information processing systems (2018)

\bibitem{muller2022instant}
M{\"u}ller, T., Evans, A., Schied, C., Keller, A.: Instant neural graphics primitives with a multiresolution hash encoding. ACM Transactions on Graphics (ToG)  \textbf{41}(4),  1--15 (2022)

\bibitem{nguyen2023single}
Nguyen, Q.H., Beksi, W.J.: Single image super-resolution via a dual interactive implicit neural network. In: Proceedings of the IEEE/CVF Winter Conference on Applications of Computer Vision. pp. 4936--4945 (2023)

\bibitem{quan2023single}
Quan, Y., Yao, X., Ji, H.: Single image defocus deblurring via implicit neural inverse kernels. In: Proceedings of the IEEE/CVF International Conference on Computer Vision. pp. 12600--12610 (2023)

\bibitem{ramasinghe2022beyond}
Ramasinghe, S., Lucey, S.: Beyond periodicity: Towards a unifying framework for activations in coordinate-mlps. In: European Conference on Computer Vision. pp. 142--158. Springer (2022)

\bibitem{ryder2022split}
Ryder, T., Zhang, C., Kang, N., Zhang, S.: Split hierarchical variational compression. In: Proceedings of the IEEE/CVF Conference on Computer Vision and Pattern Recognition. pp. 386--395 (2022)

\bibitem{saragadam2023wire}
Saragadam, V., LeJeune, D., Tan, J., Balakrishnan, G., Veeraraghavan, A., Baraniuk, R.G.: Wire: Wavelet implicit neural representations. In: Proceedings of the IEEE/CVF Conference on Computer Vision and Pattern Recognition. pp. 18507--18516 (2023)

\bibitem{sitzmann2020implicit}
Sitzmann, V., Martel, J., Bergman, A., Lindell, D., Wetzstein, G.: Implicit neural representations with periodic activation functions. Advances in neural information processing systems  \textbf{33},  7462--7473 (2020)

\bibitem{skodras2001jpeg}
Skodras, A., Christopoulos, C., Ebrahimi, T.: The jpeg 2000 still image compression standard. IEEE Signal processing magazine  \textbf{18}(5),  36--58 (2001)

\bibitem{stanley2007compositional}
Stanley, K.O.: Compositional pattern producing networks: A novel abstraction of development. Genetic programming and evolvable machines  \textbf{8},  131--162 (2007)

\bibitem{strumpler2022implicit}
Str{\"u}mpler, Y., Postels, J., Yang, R., Gool, L.V., Tombari, F.: Implicit neural representations for image compression. In: European Conference on Computer Vision. pp. 74--91. Springer (2022)

\bibitem{takikawa2021neural}
Takikawa, T., Litalien, J., Yin, K., Kreis, K., Loop, C., Nowrouzezahrai, D., Jacobson, A., McGuire, M., Fidler, S.: Neural geometric level of detail: Real-time rendering with implicit 3d shapes. In: Proceedings of the IEEE/CVF Conference on Computer Vision and Pattern Recognition. pp. 11358--11367 (2021)

\bibitem{tancik2020fourier}
Tancik, M., Srinivasan, P., Mildenhall, B., Fridovich-Keil, S., Raghavan, N., Singhal, U., Ramamoorthi, R., Barron, J., Ng, R.: Fourier features let networks learn high frequency functions in low dimensional domains. Advances in Neural Information Processing Systems  \textbf{33},  7537--7547 (2020)

\bibitem{townsend2019practical}
Townsend, J., Bird, T., Barber, D.: Practical lossless compression with latent variables using bits back coding. arXiv preprint arXiv:1901.04866  (2019)

\bibitem{wallace1991jpeg}
Wallace, G.K.: The jpeg still picture compression standard. Communications of the ACM  \textbf{34}(4),  30--44 (1991)

\bibitem{wang2003multiscale}
Wang, Z., Simoncelli, E.P., Bovik, A.C.: Multiscale structural similarity for image quality assessment. In: The Thrity-Seventh Asilomar Conference on Signals, Systems \& Computers, 2003. vol.~2, pp. 1398--1402. Ieee (2003)

\bibitem{wu20234d}
Wu, G., Yi, T., Fang, J., Xie, L., Zhang, X., Wei, W., Liu, W., Tian, Q., Wang, X.: 4d gaussian splatting for real-time dynamic scene rendering. arXiv preprint arXiv:2310.08528  (2023)

\bibitem{xie2022adan}
Xie, X., Zhou, P., Li, H., Lin, Z., Shuicheng, Y.: Adan: Adaptive nesterov momentum algorithm for faster optimizing deep models. In: Has it Trained Yet? NeurIPS 2022 Workshop (2022)

\bibitem{xu2022signal}
Xu, D., Wang, P., Jiang, Y., Fan, Z., Wang, Z.: Signal processing for implicit neural representations. Advances in Neural Information Processing Systems  \textbf{35},  13404--13418 (2022)

\bibitem{xu2022point}
Xu, Q., Xu, Z., Philip, J., Bi, S., Shu, Z., Sunkavalli, K., Neumann, U.: Point-nerf: Point-based neural radiance fields. In: Proceedings of the IEEE/CVF Conference on Computer Vision and Pattern Recognition. pp. 5438--5448 (2022)

\bibitem{linr2023}
Xu, W., Jiao, J.: Revisiting implicit neural representations in low-level vision. In: International Conference on Learning Representations Workshop (2023)

\bibitem{yan2023gs}
Yan, C., Qu, D., Wang, D., Xu, D., Wang, Z., Zhao, B., Li, X.: Gs-slam: Dense visual slam with 3d gaussian splatting. arXiv preprint arXiv:2311.11700  (2023)

\bibitem{yan2024street}
Yan, Y., Lin, H., Zhou, C., Wang, W., Sun, H., Zhan, K., Lang, X., Zhou, X., Peng, S.: Street gaussians for modeling dynamic urban scenes. arXiv preprint arXiv:2401.01339  (2024)

\bibitem{yang2023real}
Yang, Z., Yang, H., Pan, Z., Zhu, X., Zhang, L.: Real-time photorealistic dynamic scene representation and rendering with 4d gaussian splatting. arXiv preprint arXiv:2310.10642  (2023)

\bibitem{ye2023mathematical}
Ye, V., Kanazawa, A.: Mathematical supplement for the gsplat library. arXiv preprint arXiv:2312.02121  (2023)

\bibitem{gsplat2023}
Ye, V., Turkulainen, M., the Nerfstudio~team: gsplat, \url{https://github.com/nerfstudio-project/gsplat}

\bibitem{zeghidour2021soundstream}
Zeghidour, N., Luebs, A., Omran, A., Skoglund, J., Tagliasacchi, M.: Soundstream: An end-to-end neural audio codec. IEEE/ACM Transactions on Audio, Speech, and Language Processing  \textbf{30},  495--507 (2021)

\bibitem{zhang2024boosting}
Zhang, X., Yang, R., He, D., Ge, X., Xu, T., Wang, Y., Qin, H., Zhang, J.: Boosting neural representations for videos with a conditional decoder. In: Proceedings of the IEEE/CVF Conference on Computer Vision and Pattern Recognition (2024)

\bibitem{zhou2023drivinggaussian}
Zhou, X., Lin, Z., Shan, X., Wang, Y., Sun, D., Yang, M.H.: Drivinggaussian: Composite gaussian splatting for surrounding dynamic autonomous driving scenes. arXiv preprint arXiv:2312.07920  (2023)

\bibitem{zielonka2023drivable}
Zielonka, W., Bagautdinov, T., Saito, S., Zollh{\"o}fer, M., Thies, J., Romero, J.: Drivable 3d gaussian avatars. arXiv preprint arXiv:2311.08581  (2023)

\end{thebibliography}

\clearpage
\appendix

\section{Details of Gradient Computation}


In this section, we delineate the process of computing the gradients of a scalar loss function with respect to the input Gaussian parameters. Beginning with the gradient of the scalar loss $\mathcal{L}$ with respect to each pixel of the output image, we employ the standard chain rule to propagate the gradients backward toward the original input parameters.

\subsection{Gradients of Accumulated Rasterization}
The initial step involves back-propagating the loss gradients from a given pixel $i$ to the Gaussian that contributed to the pixel. For a Gaussian $n$ impacting pixel $i$, we aim to calculate the gradients with respect to its color $\boldsymbol{c}' \in \mathbb{R}^3$, the 2D means $\boldsymbol{\mu} \in \mathbb{R}^2$ and 2D covariance $\boldsymbol{\Sigma} \in \mathbb{R}^{2\times2}$.

For the color, we have 
\begin{gather}
    \frac{\partial C_i^k}{\partial c'^k_n} = \exp(-\sigma_n).
\end{gather}
where $k$ indicates the color channel.

For the $\sigma_n$, we have 
\begin{gather}
\frac{\partial C_i^k}{\partial \sigma_n} = -\exp(-\sigma_n).
\end{gather}

For the 2D mean, we have
\begin{gather}
\frac{\partial \sigma_n}{\partial \boldsymbol{\mu}_n} = \frac{\partial \sigma_n}{\partial \boldsymbol{d}_n} = \boldsymbol{\Sigma}_n^{-1} \boldsymbol{d}_n \in \mathbb{R}^2.
\end{gather}
where $\boldsymbol{d} \in \mathbb{R}^2$ is the displacement between the pixel center and the 2D Gaussian center.

For the 2D covariance, we have
\begin{gather}
\frac{\partial \sigma_n}{\partial \boldsymbol{\Sigma}_n} = -\frac{1}{2} \boldsymbol{\Sigma}_n^{-1} \boldsymbol{d}_n \boldsymbol{d}_n^\top \boldsymbol{\Sigma}_n^{-1} \in \mathbb{R}^{2\times2}.
\end{gather}
For detailed derivation of the gradient with respect to the 2D covariance, please refer to \cite{ye2023mathematical}.









\subsection{Gradients of 2D Gaussian Formation} 

Firstly, for the Cholesky decomposition, we have 
\begin{gather}
\boldsymbol{\Sigma} = \boldsymbol{L} \boldsymbol{L}^\top, 
\boldsymbol{L} = \begin{bmatrix}
    l_1 &  0 \\
    l_2 & l_3
    \end{bmatrix}.
\end{gather}
Let $G=\frac{\partial \mathcal{L}}{\partial \boldsymbol{\Sigma}}=\begin{bmatrix}
    g_1 & g_2 \\
    g_2 & g_3
\end{bmatrix}$, we then derive the gradiants of $l_1$, $l_2$ and $l_3$.

For $l_1$, we have
\begin{gather}
\frac{\partial \mathcal{L}}{\partial l_1} = \langle \frac{\partial \mathcal{L}}{\partial \boldsymbol{\Sigma}}, \frac{\partial \boldsymbol{\Sigma} }{\partial l_1} \rangle 
= \begin{bmatrix}
    g_1 & g_2 \\
    g_2 & g_3
\end{bmatrix}
\begin{bmatrix}
    2 l_1 & l_2 \\
    l_2 & 0
\end{bmatrix} 
=2g_1 l_1 + 2 g_2 l_2.
\end{gather}

For $l_2$, we have
\begin{gather}
\frac{\partial \mathcal{L}}{\partial l_2} = \langle \frac{\partial \mathcal{L}}{\partial \boldsymbol{\Sigma}}, \frac{\partial \boldsymbol{\Sigma}}{\partial l_2} \rangle
= \begin{bmatrix}
    g_1 & g_2 \\
    g_2 & g_3
\end{bmatrix}
\begin{bmatrix}
    0 & l_1 \\
    l_1 & 2l_2
\end{bmatrix} 
=2g_2 l_1 + g_2 l_2.
\end{gather}

For $l_3$, we have
\begin{gather}
\frac{\partial \mathcal{L}}{\partial l_3} = \langle \frac{\partial \mathcal{L}}{\partial \boldsymbol{\Sigma}}, \frac{\partial \boldsymbol{\Sigma} }{\partial l_3} \rangle 
= \begin{bmatrix}
    g_1 & g_2 \\
    g_2 & g_3
\end{bmatrix}
\begin{bmatrix}
    0 & 0 \\
    0 & 2l_3
\end{bmatrix} 
=2g_3 l_3.
\end{gather}

Secondly, for the rotation-scaling (RS) decomposition, we have 
\begin{gather}
\boldsymbol{\Sigma} = \boldsymbol{RSS}^\top \boldsymbol{R}^\top, 
\boldsymbol{R} = \begin{bmatrix}
    \cos{\theta} &  -\sin{\theta} \\
    \sin{\theta} & \cos{\theta}
    \end{bmatrix},
\boldsymbol{S} = \begin{bmatrix}
    s_1 &  0 \\
    0 & s_2
    \end{bmatrix}.
\end{gather}

For $\theta$, we have
\begin{gather}
\frac{\partial \mathcal{L}}{\partial \theta} = \langle \frac{\partial \mathcal{L}}{\partial \boldsymbol{\Sigma}}, \frac{\partial \boldsymbol{\Sigma}}{\partial \theta} \rangle 
= \begin{bmatrix}
    g_1 & g_2 \\
    g_2 & g_3
\end{bmatrix}
(\frac{\partial \boldsymbol{R}}{\partial \theta}\boldsymbol{S S}^\top \boldsymbol{R}^\top + \boldsymbol{R S S}^\top \frac{\partial \boldsymbol{R}^\top}{\partial \theta}).
\end{gather}
where
\begin{gather}
\frac{\partial \boldsymbol{R}}{\partial \theta}
= \begin{bmatrix}
    -\sin{\theta} &  -\cos{\theta} \\
    \cos{\theta} & -\sin{\theta}
\end{bmatrix},
\frac{\partial \boldsymbol{R}^\top}{\partial \theta}
= \begin{bmatrix}
    -\sin{\theta} &  \cos{\theta} \\
    -\cos{\theta} & -\sin{\theta}
\end{bmatrix}.
\end{gather}

For $s_1$, we have
\begin{gather}
\frac{\partial \mathcal{L}}{\partial s_1} = \frac{\partial \mathcal{L}}{\partial s_1^2} \frac{\partial s_1^2}{\partial s_1}
= \frac{\partial \mathcal{L}}{\partial s_1^2} 2 s_1
= \boldsymbol{R} \begin{bmatrix}
    2 s_1 & 0 \\
    0 & 0
\end{bmatrix} \boldsymbol{R}^\top.
\end{gather}

For $s_2$, we have
\begin{gather}
\frac{\partial \mathcal{L}}{\partial s_2} = \frac{\partial \mathcal{L}}{\partial s_2^2} \frac{\partial s_2^2}{\partial s_2}
= \frac{\partial \mathcal{L}}{\partial s_1^2} 2 s_2
= \boldsymbol{R} \begin{bmatrix}
    0 & 0 \\
    0 & 2 s_2 
\end{bmatrix} \boldsymbol{R}^\top.
\end{gather}

\section{Details of Partial Bits-Back Coding}

In Fig. 4 of Section 4.2, we show the results of our codec ("Ours"), alone with two variants using bits-back coding ("Ours+BB", "Ours-Bound"). The "Ours" is the original GaussianImage codec without any bits-back coding. It is the practical codec that achieves 2000 FPS. The "Ours+BB" is the partial bits-back coding codec described in Section 3.3. It reduces a bitrate of 
\begin{gather}
    \log (N-K)! - \log (N-K)
\end{gather}
from the original GaussianImage codec. And $K$ is selected as the lowerbound of previous $K$ 2D Gaussian whose cumulative bitrate is at least $\log(N-K)!$:
\begin{gather}
    K^* = \inf K, \textrm{s.t.} \sum_{k=1}^K R_k \ge \log (N-K)!.
\end{gather}
This rate reduction is introduced in \cite{kunze2024entropy}, and can be implemented using a first in last out entropy coder named Asymmetric Numeral Systems (ANS) \cite{Duda2013AsymmetricNS}. The encoding procedure has a sub-procedure of decoding, and the decoding procedure has a sub-procedure of encoding \cite{townsend2019practical}. The general process of partial bits-back coding is described in Algorithm~\ref{alg:lgd1} and \ref{alg:lgd2}, where $\mathcal{U}_{d}$ is uniform distribution with $d$ elements.

When applied to a whole dataset, the partial bits-back coding becomes unnecessary. More specifically, we no longer require encoding previous $K$ Gaussian for initial bits. Instead, we can just use previous image as initial bits. In that case, we can just follow the vanilla bits-back coding in \cite{townsend2019practical}. For a dataset with infinite images, the average rate reduction is
\begin{gather}
    \log N! - \log N.
\end{gather}
However, this rate reduction is never achieved as dataset is never infinite. While it is indeed the greatest lowerbound as any rate greater is achievable. Or to say, the greatest lowerbound of rate for bits-back coding is not achievable. Therefore, we name it "Ours-Bound". 

\begin{algorithm}[t]
    \caption{Partial Bits-Back Coding Encode}
    \label{alg:lgd1}
\begin{algorithmic}
    \STATE {\bfseries input} the 2D Gaussian parameters $G[1:N]$.
    \STATE {\bfseries procedure} Partial-Bits-Back-Encode($G[1:N]$)
    \STATE $m\leftarrow \emptyset$
    \FOR{$K=1$ {\bfseries to} $N$}
    \STATE $m\leftarrow$ ans-encode($m, G[K]$) \COMMENT {Rate $+R_K$ }
    \IF{len($m$) $\ge \log (N-K)!$ }
    \STATE {\bfseries break}
    \ENDIF
    \ENDFOR
    \STATE $m,G[K+1:N] \leftarrow \textrm{ans-decode}(m,\mathcal{U}_{(N-K)!},G[K+1:N])$ \COMMENT{Rate $-\log(N-K)!$}
    \STATE $m\leftarrow \textrm{ans-encode}(m,G[K+1:N])$ \COMMENT{Rate $+\sum_{i=K+1}^N R_i$}
    \STATE $m\leftarrow \textrm{ans-encode}(m,N-K)$ \COMMENT{Rate $+\log (N-K)$}
    \STATE {\bfseries return} m
\end{algorithmic}
\end{algorithm}

\begin{algorithm}[t]
    \caption{Partial Bits-Back Coding Decode}
    \label{alg:lgd2}
\begin{algorithmic}
    \STATE {\bfseries input} the bitstream $m$.
    \STATE {\bfseries procedure} Partial-Bits-Back-Decode($m$)
    \STATE $m, N-K\leftarrow \textrm{ans-decode}(m)$
    \STATE $m,G[K+1:N]\leftarrow \textrm{ans-decode}(m)$
    \STATE $m,G[K+1:N] \leftarrow \textrm{ans-encode}(m,\mathcal{U}_{(N-K)!},G[K+1:N])$
    \STATE $m,G[1:K] \leftarrow \textrm{ans-decode}(m)$
    \STATE {\bfseries return} G[1:K]
\end{algorithmic}
\end{algorithm}
\section{Experiments}
\subsection{Implementation Details}
\noindent \textbf{GaussianImage.}
During the formation of 2D Gaussian, we apply the tanh function to limit the range of position parameters to (-1,1). For covariance parameters, we add 0.5 to the diagonal elements $l_1, l_3$ of the lower triangular matrix $\boldsymbol{L}$ in the Cholesky factorization or the scaling elements $s_1, s_2$ in the rotational-scaling factorization. This adjustment prevents the scaling of the covariance from becoming excessively small. In addition, the covariance parameters and weighted color coefficients are initialized using a uniform distribution. The position parameters are initialized as follows:
\begin{equation}
  \begin{aligned}
    \boldsymbol{\mu} = {\rm atanh}({\rm rand}(2)*2-1).
    \label{eq:cholesky}
  \end{aligned}
\end{equation}
where ${\rm rand}(n)$ generates $n$ random numbers from a uniform distribution. 

\noindent \textbf{Baselines.} For SIREN \cite{sitzmann2020implicit} and WIRE \cite{saragadam2023wire}, we implement them by using the open-source project\footnote{\url{https://github.com/vishwa91/wire}} from WIRE. For I-NGP \cite{muller2022instant} and NeuRBF \cite{chen2023neurbf}, we adopt the project\footnote{\url{https://github.com/oppo-us-research/NeuRBF}} from NeuRBF. As for compression baselines, the implementation of COIN \cite{dupont2021coin} uses their library\footnote{\url{https://github.com/EmilienDupont/coin}}. We evaluate the VAE-based codecs (Ball{\'e}17 \cite{balle2017end}, Ball{\'e}18 \cite{balle2018variational}) using the MSE-optimized models provided by CompressAI \cite{begaint2020compressai}. It is worth noting that during the inference of the INR methods, we sample all image coordinates at once to output the corresponding RGB values. This is the maximum inference speed that the INR methods can achieve when the GPU memory resources are sufficient.

\subsection{Image Representation and Compression}
\begin{figure*}[t]
  \centering
  \subfloat
  {\includegraphics[scale=0.4]{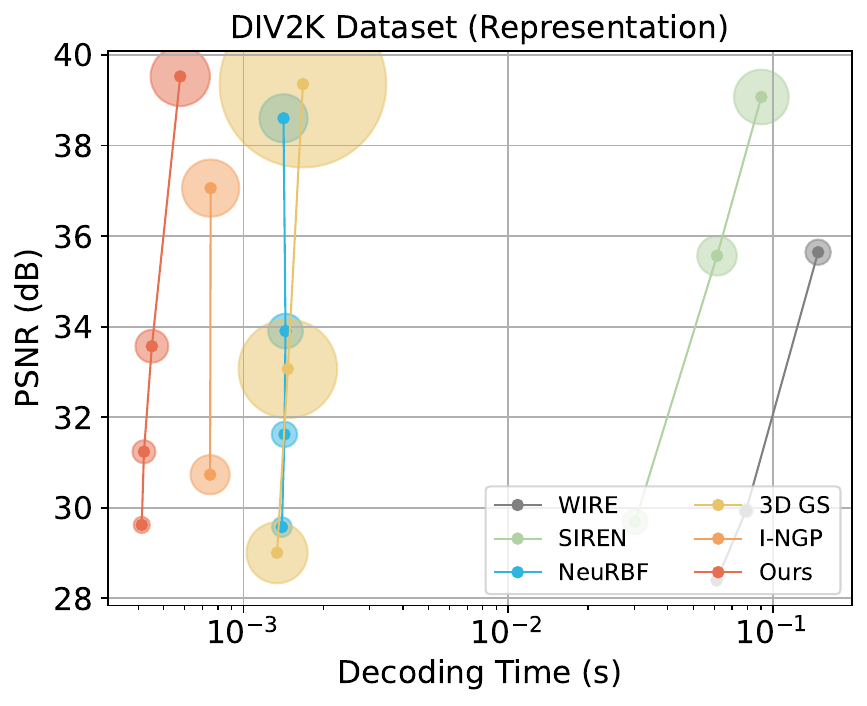}}
  \subfloat
  {\includegraphics[scale=0.4]{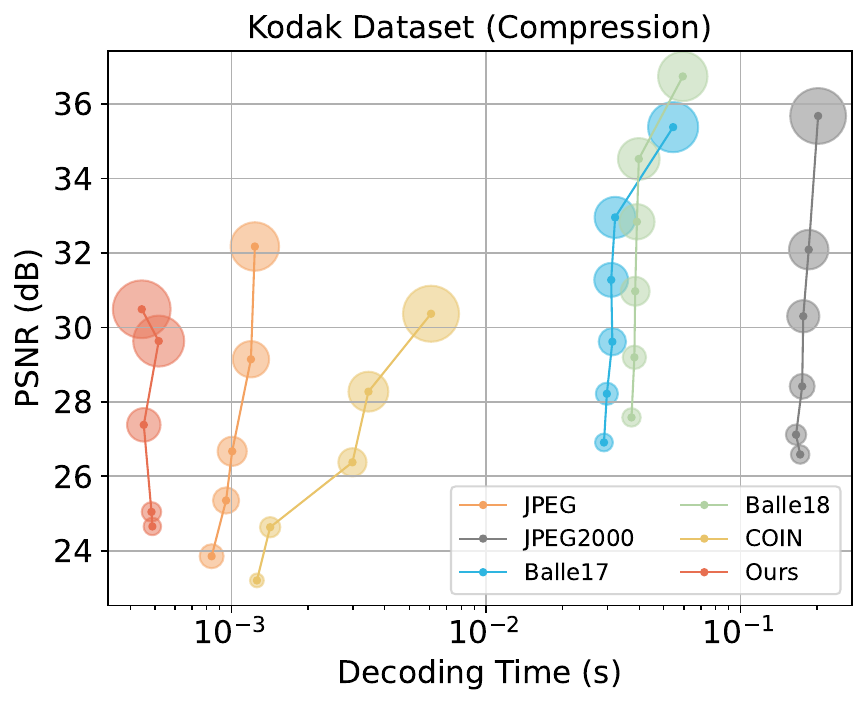}}
  \caption{Image representation (left) and compression (right) results with different decoding time on the DIV2K and Kodak dataset, respectively. The radius of each point indicates the parameter size (left) or bits per pixel (right). Our method enjoys the fastest decoding speed regardless of parameter size or bpp.}
  \label{fig:supp_fig1}
\end{figure*}

As shown in Fig.~\ref{fig:supp_fig1}, we provide performance comparisons of image representation and compression on DIV2K and Kodak datasets, respectively. 

\begin{figure*}[t]
  \centering
  \subfloat
  {\includegraphics[scale=0.39]{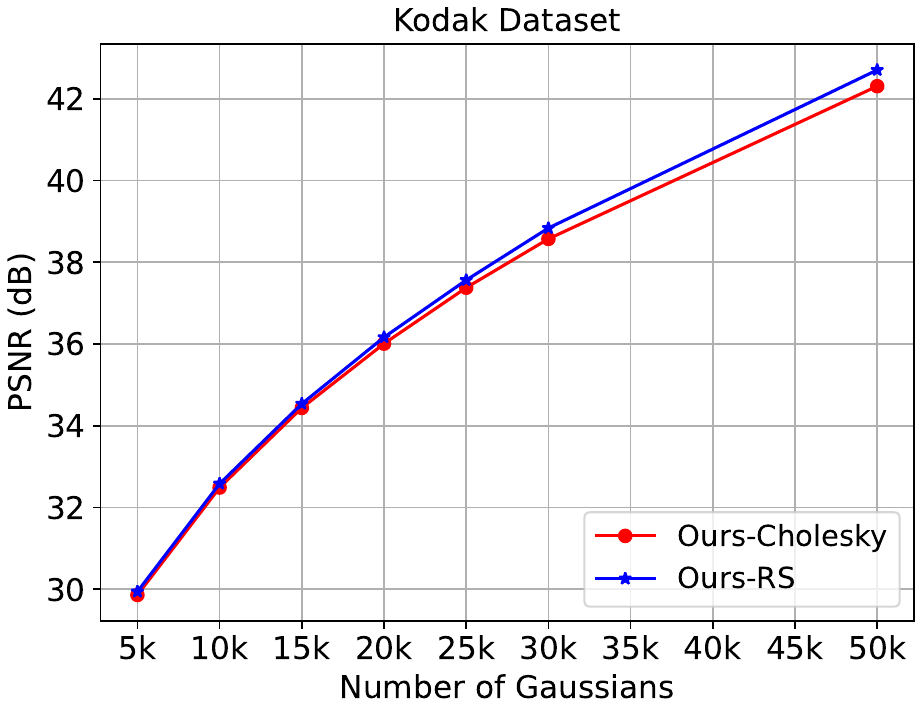}}
  \subfloat
  {\includegraphics[scale=0.39]{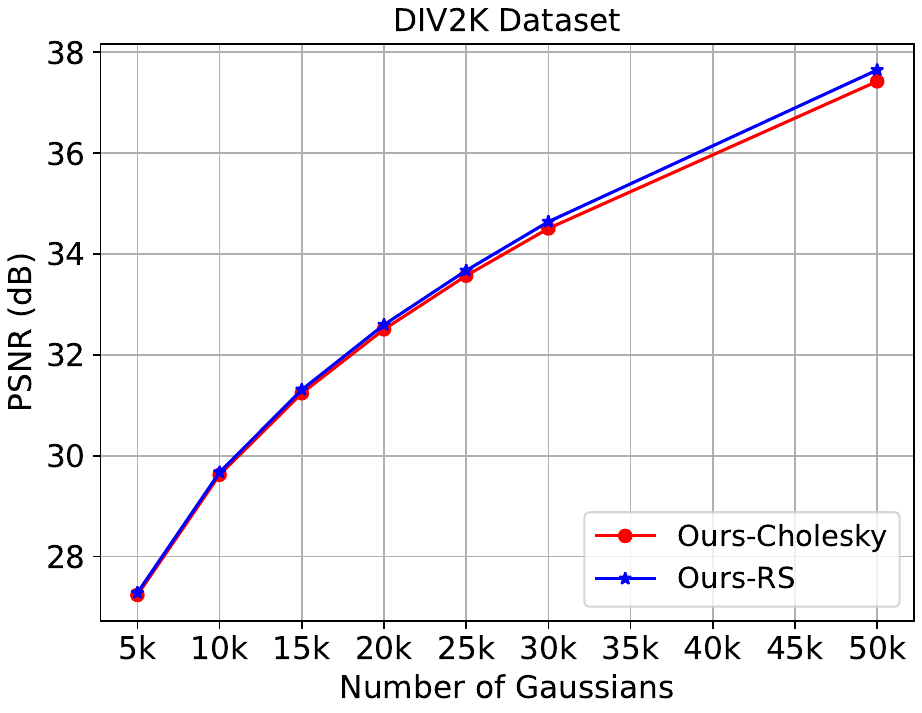}}
  \caption{Image representation with different number of Gaussians.}
  \label{fig:regression_number}
\end{figure*}

\subsection{Ablation Study}
\textbf{Number of Gaussians.} As shown in Fig.~\ref{fig:regression_number}, our proposed methods improve the quality of the fitted image as the number of Gaussians increases.

\textbf{Effect of additive operation.} The additive operation can be seen as convolving the covariance matrix with a continuous Gaussian, which helps anti-aliasing, thereby effectively improving the fitting performance, as illustrated in \tableautorefname~\ref{table:ablation_bound}.

\begin{figure*}[t]
  \centering
  \subfloat
  {\includegraphics[scale=0.41]{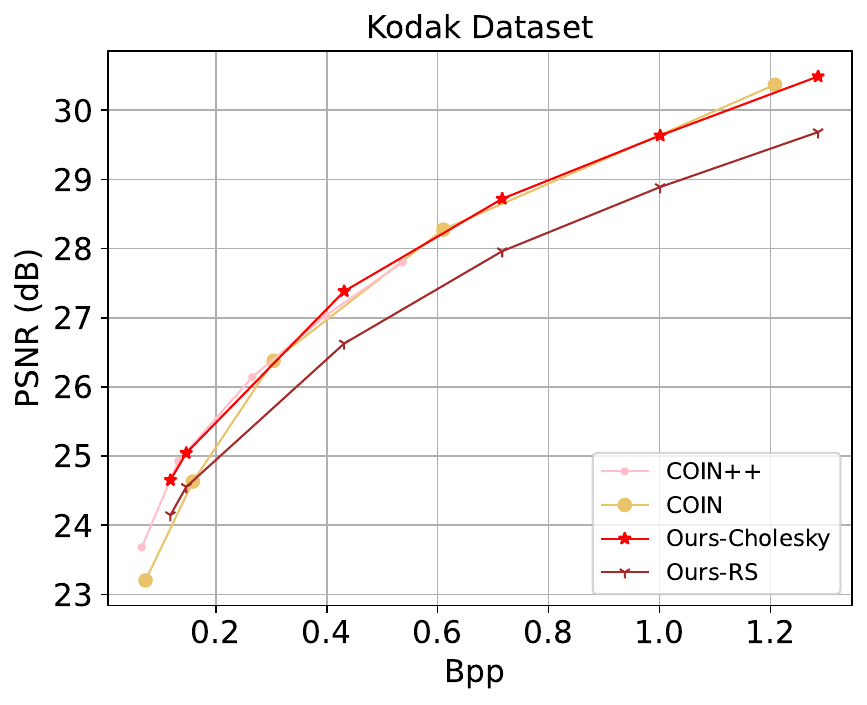}}
  \subfloat
  {\includegraphics[scale=0.41]{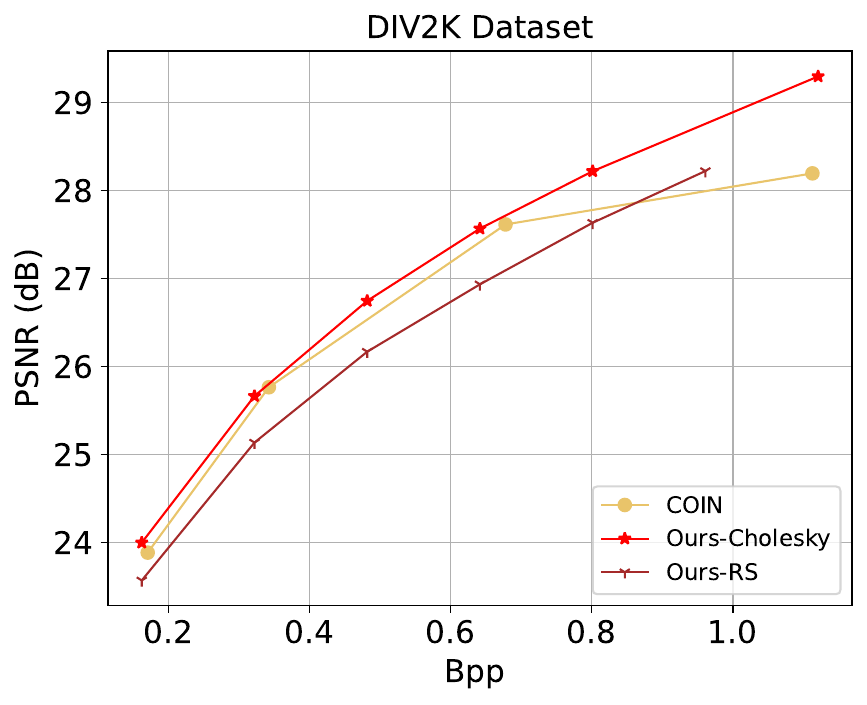}}
  \caption{Image compression results with different factorized forms on the Kodak and DIV2K dataset, respectively.}
  \label{fig:supp_fig2}
\end{figure*}

\noindent \textbf{Robustness in different factorized forms.} Fig.~\ref{fig:supp_fig2} highlights the application of identical quantization approaches to various factorized forms. Notably, the RS codec variant underperforms the Cholesky codec variant, suggesting that rotation and scaling parameters are particularly susceptible to compression distortions, which requires carefully tailored specialized quantization strategies to achieve efficient compression.

\begin{table}[t]
  \caption{Ablation study of additive operation on Kodak and DIV2K datasets with 30000 Gaussian points over 50000 training steps.}
  \label{table:ablation_bound}
  \centering
  \begin{tabular}{lcccc}
    \toprule
   \multirow{2}*{Variants} & \multicolumn{2}{c}{Kodak} & \multicolumn{2}{c}{DIV2K} \\
    \cmidrule(lr){2-3} \cmidrule(lr){4-5}& PSNR & MS-SSIM & PSNR & MS-SSIM\\
    \midrule
   Ours-Cholesky + w/ add 0.5 & 38.57 & 0.9961  & 34.51 & 0.9924  \\
   Ours-Cholesky + w/o add 0.5  & 35.05 & 0.9906 & 31.63 & 0.9830 \\
   \midrule
   Ours-RS + w/ add 0.5 & 38.83 & 0.9964 & 34.64 & 0.9927 \\
   Ours-RS + w/o add 0.5  & 36.34 & 0.9930 & 32.51 & 0.9859 \\
    \bottomrule
  \end{tabular}
\end{table}

\section{Discussion}
\label{sec:discussion}
In this paper, we simply apply existing compression techniques to build our image codec. As depicted in Fig.~4 in the main paper, there remains a considerable discrepancy between our codec and existing traditional/VAE-based codecs in the compression performance. This gap indicates an imperative need for the development of specialized compression algorithms tailored for Gaussian-based codecs to elevate the performance.
Moreover, as shown in \tableautorefname~2 in the main paper, although our encoding speed has been improved by three orders of magnitude compared with COIN \cite{dupont2021coin}, there is still a gap of four orders of magnitude compared with VAE-based codecs \cite{balle2017end, balle2018variational}. Therefore, exploring avenues for more rapid image fitting and Gaussian compression emerges as a critical research direction.

Considering that our GaussianImage provides an explicit representation and coding of images, we will further investigate this line from various aspects in the future. First, recent literature successfully performs segmentation-based text-guided editing on 3D scene represented by Gaussians~\cite{fang2023gaussianeditor,chen2024gaussianeditor}, since this discrete representation naturally provides a semantics layout. Intuitively, a similar property also exists in 2D Gaussian images, and it has the potential to develop few-shot text-guided editing on them. Second, image coding for machine~\cite{hu2020towards-icm} is a popular topic in the learned image coding community. This explicit representation is also likely to benefit downstream vision tasks like classification and detection. Finally, high-fidelity image representation~\cite{mentzer2020hific,he2022poelic} is also an essential task to delve into.


\end{document}